\documentclass[journal]{IEEEtran}
\usepackage{amsmath,amsfonts,amsthm}
\usepackage{algorithmic}
\usepackage{algorithm}
\usepackage{array}
\usepackage{textcomp}
\usepackage{stfloats}
\usepackage{url}
\usepackage{verbatim}
\usepackage{graphicx}
\usepackage{cite}
\usepackage{enumitem}
\usepackage[dvipsnames]{xcolor}
\usepackage[caption=false]{subfig}
\usepackage{booktabs}
\usepackage{multicol}
\usepackage{tabularx,ragged2e,siunitx}
\usepackage{makecell}
\usepackage[switch]{lineno} 

\usepackage[hidelinks]{hyperref}
\hypersetup{
    colorlinks,
    linkcolor={blue!50!black},
    citecolor={blue!50!black},
    urlcolor={blue!80!black}
}
\usepackage[capitalise,noabbrev]{cleveref}
\usepackage{multirow}
\usepackage{pifont}

\begin{document}

\title{Decomposition and Quantification of SOTIF Requirements for Perception Systems of Autonomous Vehicles}

\author{Ruilin Yu, Cheng Wang, Yuxin Zhang, and Fuming Zhao
\thanks{Ruilin Yu and Yuxin Zhang are with the National Key Laboratory of Automotive Chassis Integration and Bionics, Jilin University, Changchun, China. Email: \{yuxinzhang\}@jlu.edu.cn.}
\thanks{Cheng Wang is with the School of Engineering and Physical Sciences, Heriot-Watt University, Edinburgh, United Kingdom.}%
\thanks{Fuming Zhao is with Zhuoyu Technology Co., Ltd, Shenzhen, China.}%
\thanks{First Author and Second Author contributed equally to this work.}%
}



\maketitle

\begin{abstract}
Ensuring the safety of autonomous vehicles (AVs) is paramount before they can be introduced to the market. More specifically, securing the Safety of the Intended Functionality (SOTIF) poses a notable challenge; while ISO 21448 outlines numerous activities to refine the performance of AVs, it offers minimal quantitative guidance. This paper endeavors to decompose the acceptance criterion into quantitative perception requirements, aiming to furnish developers with requirements that are not only understandable but also actionable. This paper introduces a risk decomposition methodology to derive SOTIF requirements for perception. More explicitly, for subsystem-level safety requirements, we define a collision severity model to establish requirements for state uncertainty and present a Bayesian model to discern requirements for existence uncertainty. For component-level safety requirements, we proposed a decomposition method based on the Shapley value. Our findings indicate that these methods can effectively decompose the system-level safety requirements into quantitative perception requirements, potentially facilitating the safety verification of various AV components.
\end{abstract}

\section{Introduction}\label{intro}

Autonomous vehicles (AVs) have drawn much attention from the public. In addition to traditional automotive manufacturers and suppliers, many IT giants participate in this trend. As the commercialization of AVs continues to advance, safety-related issues – including methods for testing and validation of safety performance, quantitative evaluation, and improvement - have increasingly emerged as significant obstacles to their successful deployment. In principle, AVs should be at least as safe as human drivers \cite{wang2024safety}. To prove this, millions of test kilometers would be required  \cite{kalra_driving_2016}, which is infeasible for time and economic reasons. This motivates the scenario-based testing method  \cite{riedmaier2020survey}, which involves assessing system behavior using predefined and well-characterized test scenarios. 

Scenario testing has significantly reduced testing costs, but it remains prohibitively expensive to conduct tests freely. To minimize development costs, we need to establish and achieve more detailed safety goals during the development phase to ensure that AVs can pass scenario testing as much as possible. However, considering only the hazards caused by the malfunction of electrical/electronic systems is insufficient to ensure AVs’ safety since performance limitation (PL) is also a significant cause of hazards. For instance, an undetected pedestrian crossing a road could lead to an accident. Therefore, the standard ISO 21448 ``Safety of the Intended Functionality" (SOTIF) \cite{ISO21448} was proposed. In ISO 21448, safety activities are conducted to ensure there is no unreasonable risk from hazards arising from PLs, which refer to the inadequacies of the system's functional design or implementation in handling specific usage scenarios. Additionally, methodologies for analyzing PL and triggering conditions (TCs) of AV systems are presented. 

\begin{figure}[t]
	\centering
        \includegraphics[width=0.45\textwidth]{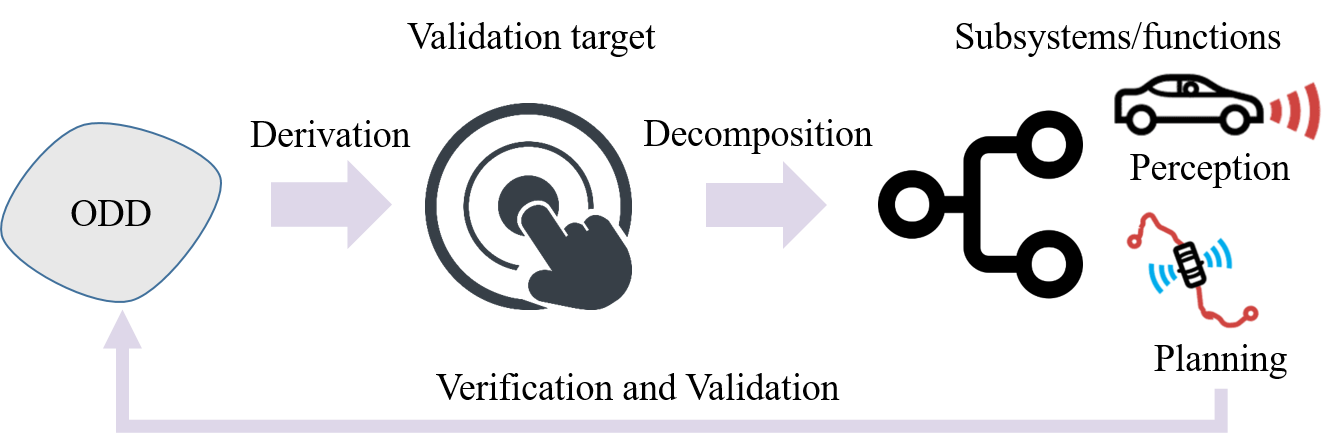}
	\caption{Requirements decomposition: a validation target is derived from a predefined Operational Design Domain (ODD) and then decomposed to requirements for subsystems or functions, which are finally verified and validated in the ODD. }
	\label{FIG:0}
\end{figure}

However, ISO 21448 lacks a systematic approach for deriving clear, comprehensible, and actionable SOTIF requirements that can be directly applied during system development. Consequently, it is unclear how good the system performance shall be, which poses challenges to the verification process  \cite{wong2020mapping}. Therefore, we discuss how to derive understandable and executable SOTIF requirements from a given acceptance criterion in the paper, as shown in \cref{FIG:0}. 

A definition of the ODD \cite{committee_taxonomy_2021} for AVs is performed during the concept phase, where an AV is supposed to be deployed. Subsequently, an acceptance criterion considering a system-level residual risk from the perspective of SOTIF is derived from the ODD. Finally, the derived acceptance criterion is decomposed into subsystems or components, formalizing their design requirements, i.e., it is necessary to define what SOTIF requirements an AV shall meet in an ODD to ensure that the residual risk is low enough.

Due to the absence of a specific method for requirement decomposition in ISO 21448, Putze et al. \cite{10186795} proposed a more general causal chain to address this issue. They argued that ISO 21448 incorrectly utilized conditional probabilities and relied on system-independent probability assumptions, which their approach aims to overcome. But, similar to ISO 21448, how the different probabilities can be obtained is not presented. Although Oboril et al.\cite{oboril_mtbf_2022} analyzed the probabilities concretely using real-world data, the probabilities consider only certain perception errors. 

Therefore, we aim to define a decomposition model in the paper to derive quantitative SOTIF requirements for perception. Specifically, we divide the safety requirements of AVs into subsystem-level and component-level. For subsystem-level safety requirements, we use Bayesian models and collision severity models to derive the requirements for the AV perception system, considering different PLs within the ODD. For component-level safety requirements, we propose the Shapley values-based method. Thus, our contributions to the paper are as follows:
\begin{itemize}
    \item A generic model to decompose the acceptance criterion into AV perception requirements is proposed, considering both object detection error rate and accuracy. 
    \item Different intended behavior models are analyzed and compared to select an appropriate model as a basis for deriving perception requirements; 
    \item A model-independent method based on Shapley value is proposed to quantify the safety requirements of AV components;
\end{itemize}

The paper is organized as follows. We introduce related work in \cref{rela}. \cref{meth} presents the methodology we applied to decompose the acceptance criterion to subsystem-level and component-level. \cref{result} describes the implementation of the proposed approach and demonstration of the derived requirements. Finally, the discussion and conclusion are shown in \cref{disc} and \cref{con}, respectively. 

\section{Related works}\label{rela}

To derive the acceptance criterion in an ODD, the principles of ``As Low As Reasonably Practicable" (ALARP), ``Global Au Moins Equivalent" (GAME) and ``Minimum Endogenous Mortality" (MEM) are applicable  \cite{junietz2019macroscopic}. ALARP refers to reducing the level of residual risk to a reasonable and feasible extent, while the GAME acceptability criterion refers to any new system or modification to an existing system that should be safer than the recognized reference systems. MEM, on the other hand, is a method of deriving absolute risk acceptance values based on natural mortality rates for specific age groups \cite{en_50126-2_2007}. 

AVs, as a new technology, lack existing reference systems and an explicitly defined reasonably practicable residual risk. A common approach to address this challenge is to use skilled and experienced human driver performance as a benchmark \cite{un_ece_regulation_2021, wang2023application}. As the Ethics Commission on Automated and Connected Driving of the German Federal Ministry of Transport stated: ``The licensing of automated systems is not justifiable unless it promises to produce at least a diminution in harm compared with human driving."\cite{federal_2017}. This concept is also referred to as a Positive Risk Balance in some studies  \cite{kauffmann2022positive, fahrenkrog2023implications}. Berk\cite{berk_safety_2019} calculated this value, as approximately $\rm 6.8 \times 10^{-7}/h$ based on US accident rate and $\rm 1.5 \times 10^{-7}/h$ based on German highway accident rate. This value will be used as an input for the subsequent decomposition of the safety requirements. Since the focus of this paper is on SOTIF, safety requirements for other aspects of the AV are not considered in this paper. In the following, it is referred to as the system-level safety requirement.

Regarding system-level risk decomposition, Oboril et al.\cite{oboril_mtbf_2022} introduced a mean time between failure model to establish the relationship between perception errors and system-level risks. They concluded that if the system-level risk is $10^{-5}$ failures/h, the false negative (FN) rate should be less than $5.0 \times 10^{-5}$ failures/h. They did not consider errors from the planning module since responsibility-sensitive safety (RSS)  \cite{shalev-shwartz_formal_2017} could ensure safe operation. The quantified requirements from this model are then used as verification guidelines for corresponding perception algorithms, e.g., multi-object tracking (MOT). However, in addition to the FN rate, other SOTIF requirements, such as the false positive (FP) rate and velocity accuracy, are also essential but were neglected in their model.  

Buerkle et al. \cite{buerkle_safe_2021} used the RSS model to establish the relationship between position errors and collision severity, determining the maximum allowed position error for a tracked object based on acceptable collision severity. However, they did not present the relationship between position errors and system-level risk. Beyond position errors, the detection range is also a critical requirement for perception systems. Ye et al. \cite{ye_operational_2022} examined the necessary detection range and angle for AV sensors under various ramp and road speed limits. Berk \cite{berk_safety_2019} proposed a method to break down perception requirements into sensor specifications by considering sensor fusion algorithms. A perception requirement with a $10^{-9}$ FP and FN rate per hour was used as a case study to derive acceptable sensor error rates. These derived sensor error rates can then be utilized to select appropriate sensors that meet the desired perception requirements. Clearly, perception requirements are foundational to determining sensor specifications. Chu et al. \cite{chu_sotif-oriented_2023} identified the minimum required perception area and error rates for a specific driving scenario using a safety distance model. Qiu et al. \cite{qiu_reliability_2021} focused on reliability analysis for a multi-sensor system by calculating the system error rate, which depends on FP and FN errors across different fields of view. They applied a Markov-based approach to model sensor correlations and compared system error rates across four different sensor fusion strategies, both with and without considering sensor correlation. Their findings highlighted the importance of sensor correlation in reliability analysis and suggested that the derived system error rate  is valuable for quantitative SOTIF analysis.

Although numerous studies have proposed methods to address the challenge of specifying SOTIF requirements for AV perception systems, many fail to establish a clear connection between SOTIF requirements and the acceptance criterion, making it difficult to define precise specifications. Notably, Oboril \cite{oboril_mtbf_2022} introduced a method aimed at addressing this limitation. However, this approach only links the FN error rate to the acceptance criterion, neglecting other types of perception errors. While Oboril's method shows promise in specifying the FN error rate, it requires further refinement for broader application. In addition, while some studies have considered the decomposition of component-level safety requirements, their approaches have been conducted based on specific architectures. For example, Berk \cite{berk_safety_2019} used a voting machine model to decompose perception safety requirements, and Qiu et al. \cite{qiu_reliability_2021} used a Markov model to simulate the perception process. To date, no generalized methodology comprehensively quantifies both subsystem-level and component-level safety requirements for SOTIF in AVs. To address this gap, we first differentiate between subsystem-level and component-level safety requirements. For subsystem-level safety, we propose a Bayesian model and a collision severity model, and for component-level safety, we introduce a Shapley value-based decomposition method.

\section{Methodology}\label{meth}

\begin{figure*}[t]
	\centering
        \includegraphics[width=0.95\textwidth]{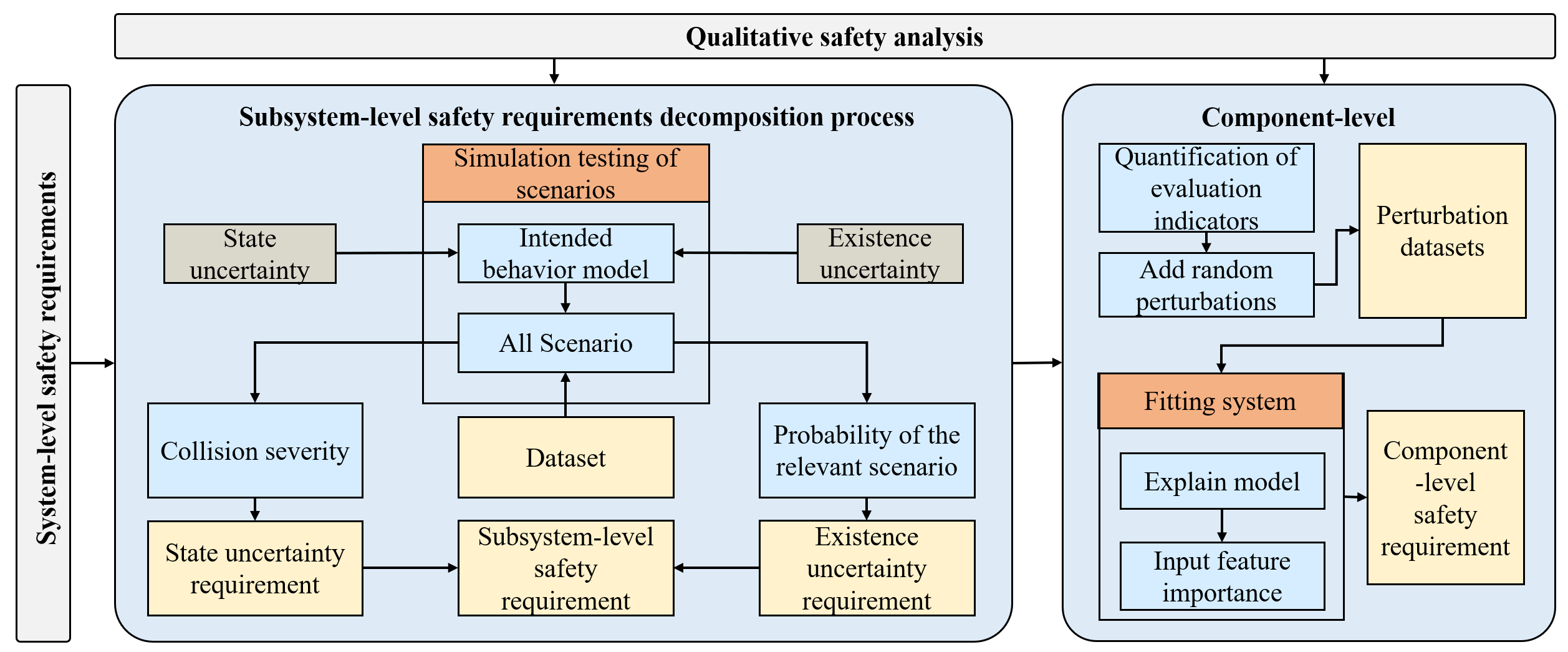}
	\caption{The implementation process of the proposed methodology to derive SOTIF requirements.}
	\label{FIG:2}
\end{figure*}

The proposed method is illustrated in \cref{FIG:2}. System-level safety requirements are first decomposed into subsystem-level safety requirements, which are subsequently broken down into component-level safety requirements. 


The subsystem-level decomposition process begins with system-level safety requirements, with the aim of ensuring that the entire system adheres to these standards. Subsequently, we select the appropriate intended behavior model as the basis for deriving perception safety indicators. Then, depending on the different performance limitations (PLs) in the ODD, we use a Bayesian model and a collision severity model to derive the safety requirements for the AV sensing system.

At the component level, the focus shifts to individual system components and their specific safety requirements. This stage involves quantifying evaluation metrics, introducing random perturbations to assess robustness, and using perturbation datasets to challenge the system. The system response is then analyzed by fitting models and interpreting their behavior, particularly with regard to the importance of input features. Given the significant architectural and functional differences among AV components, we propose introducing the Kernel Shapley Additive Explanations method—a model-agnostic approach—to quantify safety requirements for these components.

\subsection{Modeling of Risk Decomposition}

Since the quantification method will be derived from the conceptual framework of the standard discussed in the following sections, several key concepts need to be clarified to avoid misunderstandings. The definitions of these concepts are as follows:

\begin{itemize}
  \item \textbf{Harm}: Physical injury or damage to persons, property, or livestock.
  \item \textbf{Risk}: Combination of the probability of occurrence of harm and the severity of that harm
  \item \textbf{Hazard}: A potential source of harm arising from hazardous behavior at the vehicle level.
  \item \textbf{Hazardous event}: An event that has the potential to cause harm.
  \item \textbf{Hazardous behavior}: Actions or behaviors caused by limitations or defects in a system, function, or algorithm, which result in the system failing to perform its tasks safely or as expected, potentially leading to harm or danger.
\end{itemize}

\begin{figure}[t]
	\centering
        \includegraphics[width=0.5\textwidth]{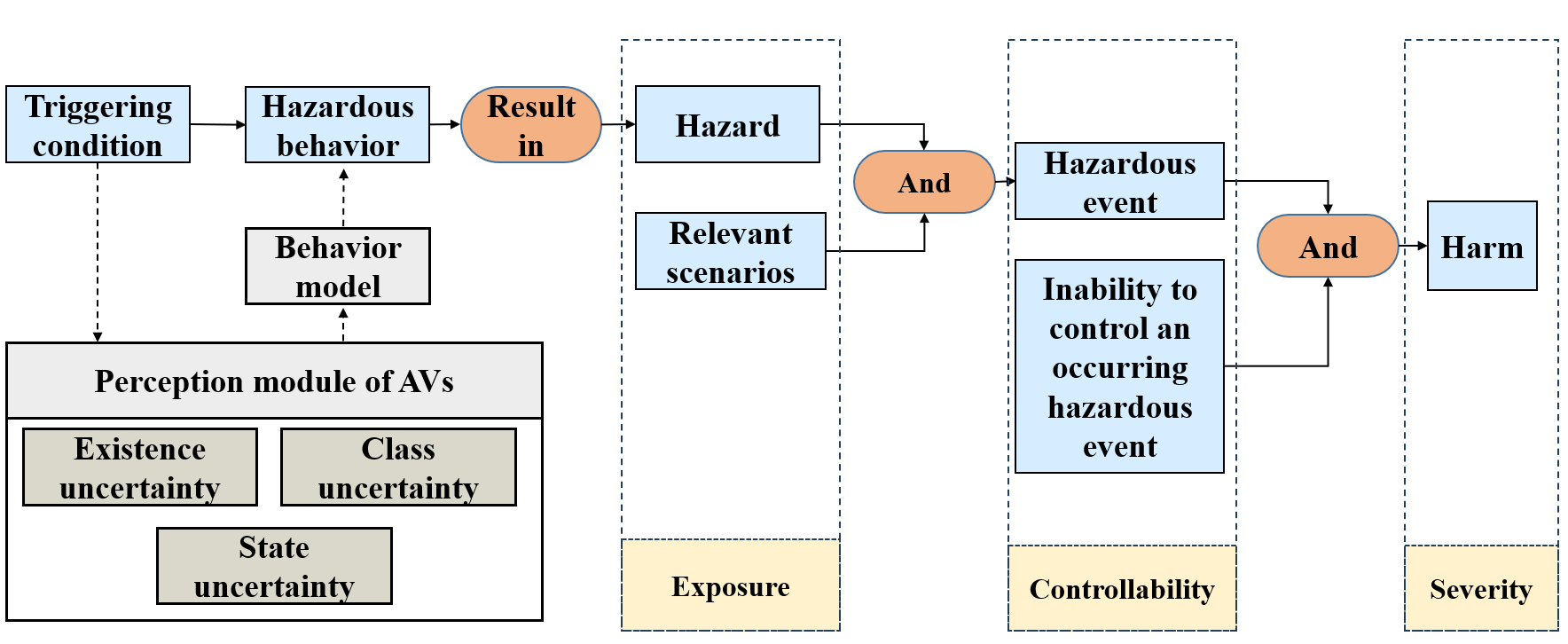}
	\caption{The concept of decomposing system-level risk into SOTIF requirements for perception.}
	\label{FIG:1}
\end{figure}

According to ISO 21448, harm occurs when a hazardous event cannot be controlled, and a hazardous event arises from the combination of a hazard and a relevant scenario. For example, a rapid deceleration is a hazard that can lead to a hazardous event if a vehicle is closely following. Moreover, a hazard results from hazardous behavior, such as rapid deceleration caused by an FN object, with a triggering condition as the cause of such behavior, as illustrated in  \cref{FIG:1}. From the figure, it can be seen that ISO 21448 describes the process of harm occurrence from three perspectives: exposure, controllability, and severity.

Focusing specifically on AV perception systems, TC can impact the performance of the system, potentially leading to three types of uncertainty
\cite{dietmayer_pradiktion_2015}: 
\begin{itemize}
  \item \textbf{Existence uncertainty }: uncertainty on whether existing objects are detected and whether non-existing ghost objects are wrongly indicated.
  \item \textbf{State uncertainty}: uncertainty on the semantic types of detected objects.
  \item \textbf{Class uncertainty}: uncertainty on the state of physical quantities of detected objects.
\end{itemize}

These uncertainties are categorized as PLs resulting from TC. When present, they can lead to hazardous behaviors through the implemented behavior model \cite{9765675}. For instance, consider a scenario where the TC is heavy rain. This condition might generate an FP object due to sensor noise. Consequently, the AV may brake unexpectedly because of commands from the behavior model, resulting in unintended hazardous braking. Using this model, we can establish requirements for managing uncertainties based on a predefined harm acceptance criterion.

Based on the process illustrated in \cref{FIG:1}, we derive more detailed safety requirements by tracing backward from the system-level safety requirements. The relationship between PLs and the system-level risk $\lambda$ is expressed in \cref{eq1}. In \cref{eq1}, we assume that all PLs are mutually exclusive, making their joint probability necessarily greater than that under dependency conditions. If the joint probability under the mutual exclusivity assumption can be shown to satisfy the system-level safety acceptance criterion, it follows that the joint probability under any dependency conditions will also satisfy the criterion.
\begin{equation}\begin{split}\label{eq1}
\lambda=\sum_{j}\sum_{k=1}^{\rm N}{\left(p_{{\rm PL},j,k}\times p_{{\rm E|PL},j,k}\times p_{\rm C|E}\times p_{\rm S|C}\right)}\\
{\rm with} \ j=( {\rm perceptipn,plan,actuator}), k=({ \rm 1,\cdots,N})
\end{split}\end{equation}
where $p_{{\rm PL},j,k}$ represents the probability of the $k$th PL in the $j$th module, while $p_{{\rm E|PL}}$ is the probability of a scenario that can lead to a hazardous event once the PL exists. $p_{\rm C|E}$ stands for the probability of the scenario that isn't controllable under the condition of a relevant scenario. The last term $p_{\rm S|C}$, represents the probability of a harm occurrence with a certain severity level when AV is uncontrollable. This term should be adjusted according to the given acceptance criterion, i.e., if $\lambda$ is the statistical result from fatal accidents, $p_{\rm S|C}$ will be the probability of fatal accidents occurring in situations where AV is uncontrollable. $N$ is the total number of PLs of an AV. However, the PL of the AV can vary with each iteration, meaning that $N$ continuously changes throughout the iterative process.

The proposed risk decomposition methodology aligns with several well-established safety frameworks, including Failure Modes and Effects Analysis (FMEA) and the System-Theoretic Accident Model and Processes (STAMP). These methods are typically employed during the qualitative safety analysis phase and can serve as valuable inputs to the quantitative methods proposed in this paper. Specifically, the three types of uncertainties identified in this study partially aligned with the failure modes identified through FMEA in the analysis of perception systems. Similarly, the controllability and severity assessments presented in \cref{FIG:1} integrate seamlessly with STAMP's control structure analysis, enabling the efficient identification of uncontrollable hazardous events, unsafe control actions and corresponding causal scenarios.

The overall safety of an AV is influenced by the PLs of each module, as illustrated in \cref{eq1}. Plan and actuator modules, being primarily deterministic with minimal internal uncertainty, are assumed to have a negligible impact on system-level risk. This assumption is supported by the fact that these modules can be thoroughly verified in simulations against specific requirements \cite{xu_calibration_2021}. In contrast, the perception module, which is inherently more complex and prone to uncertainties \cite{10422266}, plays a significant role in risk occurrence. Consequently, our focus is on the requirement decomposition for the perception system to address these challenges effectively.

To quantify perception requirements, we first determine the maximum allowable uncertainties that meet the acceptable levels of collision severity. This requires first extracting relevant scenarios from the dataset, such as cut-in situations. These scenarios are used to compare different available intended behavior models, ultimately selecting the most appropriate model for the collision severity analysis. Within the collision severity model, various degrees of state uncertainty, such as position errors, are introduced into the chosen behavior model. Subsequently, these behavior models, now incorporating uncertainty, are simulated within the relevant scenarios. Through simulation, the resulting collision velocity, an indicator of the severity of the potential collision, is calculated \cite{buerkle_safe_2021}. Given the maximum acceptable collision severity, we can then mathematically determine the maximum allowable state uncertainty that would still meet this criterion.

Class uncertainty is directly related to object state uncertainty because different prediction models are used for different object classes. For example, a constant velocity model may be more appropriate for predicting the behavior of a truck, while a constant turn rate and velocity model may be better suited for a car \cite{weng_vehicle_2014}. As a result, if an object is incorrectly classified, the associated prediction model may not accurately represent its behavior, leading to increased state uncertainty. Therefore, class uncertainty can be transformed into state uncertainty by injecting it into the prediction or plan model. After that, the quantification method is the same as state uncertainty, which will not be repeated in this paper.

Once the requirements for state uncertainty are established, an AV must meet these criteria during the verification phase. This ensures a significant reduction in system-level risk attributed to state uncertainty. Therefore, for the requirement of state uncertainty, the probability of state uncertainty occurrence $p_{{\rm PL},j,k}$ is no longer used as a constraint, but the maximum allowable state uncertainty error is used as a constraint. Consequently, \cref{eq1} can be revised as follows: 
\begin{equation}\begin{split}\label{eq2}
\lambda=\sum_{j}\sum_{k=1}^{\rm N}&\left(p_{{\rm PL},j,k}\times p_{{\rm E|PL},j,k}\times p_{\rm C|E}\times p_{\rm S|C}\right)\\
&{\rm with}\  j=({\rm sense}),\rm N=2
\end{split}\end{equation}
where $k$=1 represents the FN and $k$=2 is the FP. 

Using the Bayesian model described in \cref{eq2}, we begin by calculating the probability of relevant scenarios using the Aerial Dataset for China's Congested Highways and Expressways (AD4CHE) dataset\cite{zhang2023ad4che}. This process introduces uncertainty into the intended behavior model. Any collision incident observed during simulation testing is categorically identified as a relevant scenario. Next, through a detailed analysis of scenario parameters within the ODD, we identify a collection of relevant scenarios. This identification process clarifies the combinations of scenario parameters where uncertainty could lead to accidents within the selected behavior model. Based on these identified parameters, we extract the relevant scenarios and calculate their probability by dividing the number of relevant scenarios by the total number of scenarios in the dataset. Using this probability, we then derive the requirements for existence uncertainty as outlined in \cref{eq2}.

Finally, we derive the requirements for both state uncertainty and existence uncertainty. In the following subsections, we will detail each critical step involved, as illustrated in \cref{FIG:2}.

\subsection{Intended behavior models}
The subsystem-level safety requirement decomposition method involves identifying the performance boundaries of the AV planning system within a simulation environment. Once these boundaries are established, the subsystem-level safety requirements that satisfy system-level safety standards are derived by inversely mapping from the performance thresholds. 

A more robust AV planning system allows for more lenient safety requirements for the perception system. But in this study, we used the intended behavior model as a proxy for the planning system in our simulation experiments. In order to bridge the performance gap between intended behavioral model and AV planning systems, we will compare mainstream intended behavioral models in what follows and select the one with the best safety performance as the basis for subsequent analysis.

We review the intended behavior models to select the most suitable one for deriving perception requirements. One well-known model is the RSS, which is commonly used in AV planning as a safety check module to ensure safe decisions \cite{nageshrao_autonomous_2019}. Two additional models, the Competent and Careful Driver (CC Driver) and the Fuzzy Safety Model (FSM), were proposed in UN Regulation 157 \cite{un_ece_regulation_2021}.

The FSM tries to mimic the driving strategy of humans by using a fuzzy process, while the RSS model defines a conservative minimum safe distance that an AV should not fall below. The CC Driver, on the other hand, focuses more on a driver's extreme driving capability. Either no braking or full braking is applied in this model. Before we choose an appropriate intended behavior model for our requirement decomposition, we compare and analyze the three models mentioned above using the AD4CHE dataset to determine if the FSM is still the most suitable one in traffic situations in China.

\subsection{Collision Severity Model}
Based on the intended behavior model, we can derive the necessary requirements for the perception system to ensure the safe execution of the intended behavior. Specifically, if there is any state uncertainty within the perception system, the intended behavior may not be executed safely, leading to an increased risk of collision. To address this, we propose a model that quantifies the relationship between state uncertainty and collision severity. In this model, we use $\Delta v$ to represent the collision risk. Here, $\mathcal{F}\left(\zeta\right)$ denotes the expected output of the intended behavior model given a specific perception input $\zeta$, while $\mathcal{F}\left(\zeta+u\right)$ represents the actual output when there is some uncertainty $u$ in the input. Our objective is to identify the maximum allowable uncertainty that maintains $\Delta v$ within a predefined upper limit. The relationship can be mathematically expressed as:
\begin{equation}
\label{eq8}
\begin{split}
    \mathop{\arg\max}\limits_{u}\ &\Delta v\sim \mathcal{F}\left(\zeta+u\right)-\mathcal{F}\left(\zeta\right)
\\
    & s.t.\ \ \Delta v \leq \Delta v_{\rm max}  
\end{split}
\end{equation}

Finally, the derived maximum allowable uncertainty is treated as a SOTIF requirement and allocated to the appropriate module in the perception system. For example, when we consider the position uncertainty of an object, the solved results according to \cref{eq8} are regarded as the requirements for a MOT algorithm. 

\subsection{Bayesian Model}
After fulfilling the requirements derived from the collision severity model, the next challenge is addressing existence uncertainty. \cref{eq2} outlines the relationship between system-level risk and existence uncertainty. Currently, existence uncertainty within the sensor’s field of view is treated as a uniform entity, without considering variations based on relative distance. This approach can lead to requirements that are either too stringent or too lenient. A more effective strategy is to assign varying requirements for existence uncertainty according to distance, with stricter requirements in nearby areas and more relaxed ones at greater distances. This approach aligns well with the actual performance of perception systems \cite{10315197}. Thus, \cref{eq2} is extended to
\begin{flalign}\label{eq9}
\lambda=\sum_{k}\sum_{j=1}^{N}&\left(\sum_{i=1}^{M}{{(p}_{{\rm PL},i,j,k}\times p_{{\rm E|PL},i,j,k})}\times p_{\rm C|E}\times p_{\rm S|C}\right)
&\\
\nonumber &{\rm with}\ k=\left({\rm sense}\right),\ N=2
\end{flalign}
where $M$ represents the number of distance partitions. In a specific distance partition $A_i$, $p_{{\rm PL},i,j,k}$ means the probability of existence uncertainty in that partition. $p_{{\rm E|PL},i,j,k}$ can be further defined as
\begin{equation}\label{eq10}
    p_{{\rm E|PL},i,j,k}=p\left\{d(S)\le d_{\rm safe},\ \forall S\in A_i\right\}
\end{equation}
\begin{figure}[t]
	\centering
        \includegraphics[width=0.45\textwidth]{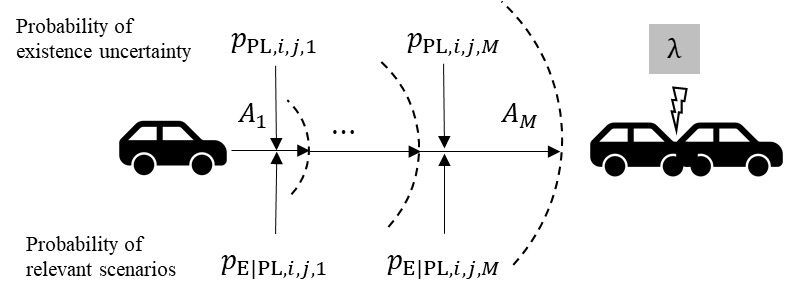}
	\caption{The concept of the distance-based model to derive the requirements for existence uncertainty in each distance partition.}
	\label{FIG:3}
\end{figure}

For each scenario in $A_i$, $p_{{\rm E|PL},i,j,k}$ represents the probability of a relevant scenario, which is identified when the relative longitudinal distance $d(\rm S)$ is less than the safe distance $d_{\rm safe}$ derived from the intended behavior model. \cref{FIG:3} illustrates the Bayesian model, quantified by \cref{eq9} and \cref{eq10}, which establishes the relationship between system-level risk $\lambda$ and existence uncertainty across the entire detection range. By utilizing this model, we can derive more realistic and appropriate requirements for the existence uncertainty in different detection zones. This method not only ensures that requirements are tailored to specific distance partitions but also facilitates independent verification, avoiding overly stringent requirements in distant partitions.

\subsection{Shapley value-based decomposition method}

Subsystem-level safety requirements, as outlined in \cref{meth}, are well-suited for the development of simple systems. For instance, in a single-sensor automatic emergency braking system, a requirement such as "no more than X failures detected within 25 to 50 meters" allows developers to iterate effectively. Simple systems are typically managed by a single team, enabling independent optimization of performance, and allowing direct assignment of subsystem-level requirements. Therefore, for simpler systems, it is sufficient to quantify security requirements at the subsystem-level. For L2 and higher-level perception systems, the aforementioned requirements will significantly increase testing and development costs. This is due to the lack of knowledge regarding how performance optimizations of individual components contribute to overall system performance. As a result, a more granular approach to the component level is necessary to improve efficiency.

To address the challenges posed by complex systems, we propose a method for decomposing component-level safety requirements using Shapley values. This approach allows for a fair and systematic allocation of safety requirements across subsystems, ensuring that each component's contribution to overall safety is appropriately quantified and addressed.

The Shapley value is a cooperative game-theoretic concept for measuring the contribution of each player\cite{shapley1953value}. It equitably distributes total rewards based on individual contributions and represents the average marginal contribution of a feature across all possible coalitions\cite{ancona2019explaining}. Although it uniquely quantifies contributions and can effectively measure feature importance, its computation is complex and scales exponentially with the number of features. To address this problem, Lundberg and Lee \cite{lundberg2017unified} simplify the process by using a weighted linear model to account for complex models. They generated a new dataset by replacing the original data with background values (e.g., 0 or the mean), trained the explanatory model on this modified dataset, and ensured its predictions aligned with those of the complex model, as described in \cref{eq:linear_model,eq:objective,eq:contraint}. 

1) a weighted linear model needs to be constructed:
\begin{equation}\label{eq:linear_model}
    g\left(z\right)=\phi_0+\sum_{j=1}^{M}{\phi_jz_j}
\end{equation}
where \( g \) denotes the interpretive model, \( z \) represents the sample data with randomly replaced, \( \phi_0 \) signifies the average value of \( f(x) \), \( z_j \) denotes the simplified feature, and \( z_j \) takes the value 1 if the corresponding original feature has been replaced, and 0 otherwise.

2) the interpretive model \( g(z) \) is trained. Based on the aforementioned criteria, we define the loss function as follows:

\begin{equation}\label{eq:objective}
   arg \, \text{min}\, L(f,g,\pi_x)= \sum_{z\in N}[f(x)-g(z)]^2\pi_x(z)
\end{equation}
\setlength{\arraycolsep}{0.7pt}
\setlength{\arraycolsep}{1pt}
\begin{equation}\label{eq:contraint}
    \pi_x(z) = \frac{M-1}{\binom{M}{|z|}|z| (M-|z|)} 
\end{equation}
where $f$ is the original prediction model to be explained, \( M \) is the total number of features of $z$, and \( |z| \) is the number of replaced features in $z$. By training, the coefficients $\phi_j$ of the regression equation that minimize $L(f, g, \pi_x)$ are obtained. By using the above weight  $\pi_x(z)$, it can be proven that the trained weighted linear model coefficients are the Shapley values of each feature\cite{lundberg2017unified}.

\begin{figure*}[htbp]

\centering

\includegraphics[width=0.90\textwidth]{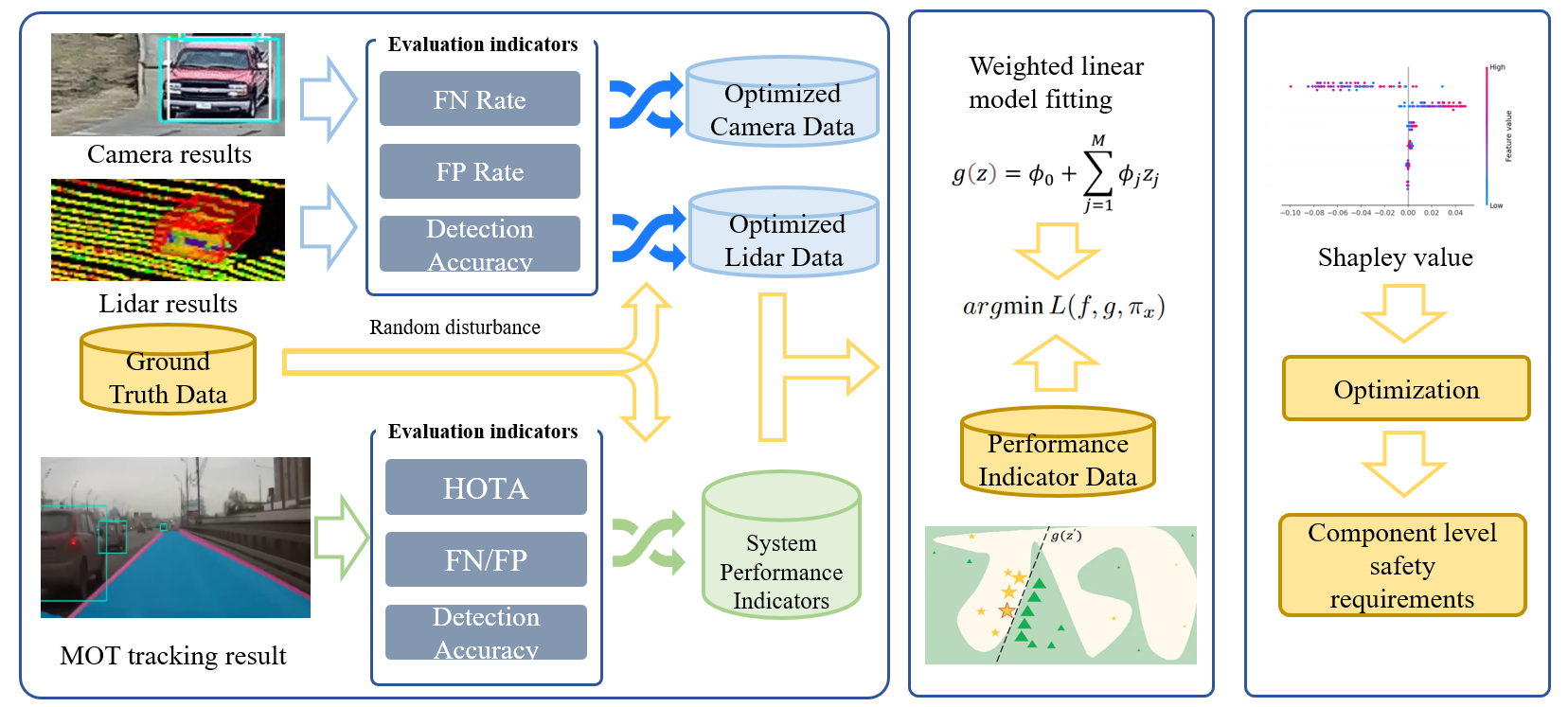}

\caption{ MOT algorithm is treated as a function characterized by predefined input and output data evaluation metrics, where numerical inputs and outputs are used to reflect changes in algorithm performance. Next, based on the variations in these input metrics, we generate a sample dataset that captures a range of possible performance scenarios. A weighted linear model is then optimized on this dataset to identify model parameters that minimize the loss function. These parameters, representing the Shapley values, quantify the influence of each input metric on the output metric. Finally, component-level safety requirements are allocated based on these calculated Shapley values.}

\label{fig:framework}

\end{figure*}

This approach is applied to decompose safety requirements for complex systems within AV. As a case study, we focus on the MOT system, which is essential for accurate perception in AVs. We decompose the safety requirements for the MOT algorithm into specific output metrics for the Lidar and camera algorithms, as detailed in \cref{fig:framework}.

To accurately decompose the subsystem-level safety requirements, it is first essential to select appropriate performance evaluation metrics for both input and output data. These metrics will be used to train the weighted linear model. The selection process should be informed by a qualitative analysis of the system under evaluation. For each subsystem-level safety requirement, a corresponding output performance metric must be chosen. Similarly, for each component-level safety requirement, an appropriate input performance metric needs to be identified.

In the case of the MOT system, three key input metrics—FN, FP, and bounding box detection accuracy—were selected to assess the detection performance of the Lidar and camera fusion system. The accuracy of bounding box detection is determined by summing the errors in the horizontal and vertical coordinates of the corners of the detected bounding box relative to the ground truth. To evaluate the output of the MOT algorithm, FN, FP, Higher Order Tracking Accuracy (HOTA), and its three sub-metrics are selected as output metrics \cite{luiten2021hota}. HOTA is a comprehensive metric that assesses MOT performance by decomposing it into detection, association, and localization tasks, with the final HOTA score reflecting a combination of these aspects.

Once these evaluation metrics are selected, we introduce controlled perturbations into the input data to simulate variations, adjusting the dataset to reflect these changes.

This perturbed data is then input into the MOT algorithm to observe the effects on the output metrics. An explanatory model is trained using the perturbed input and corresponding output performance metrics. The coefficients of this model represent the Shapley values, quantifying the impact of each input metric on each output metric, thereby guiding the allocation of component-level safety requirements.

Then,component-level safety requirements are assigned according to the following formula:
\begin{equation}
    P(s)=f(x)=g(x)=\phi_{0}+\sum_{j=1}^{M} \phi_{j} P_{j}(e)<P(r)
\end{equation}
where \( P(s) \) represents the subsystem-level failure probability, and \( P(r) \) denotes the subsystem-level safety requirement. The condition \( P(s) < P(r) \) indicates that the system meets the predefined subsystem-level safety requirements. In this study, \( P(s) \) is equal to the performance evaluation metric of the function \( f \) under the original data \( x \). Through training, it can be considered that \( f(x) = g(x) \). Subsequently, the simplified feature \( z_j \) in \cref{eq:objective} is replaced with the component-level safety requirement to be allocated \( P_j\left(e\right) \). This yields the constraints between the subsystem-level safety requirement and the component-level safety requirement.

Lastly, under these established conditions, efforts are made to minimize the optimization cost associated with each component-level requirement while effectively allocating component-level safety requirements needs to meet overarching subsystem-level safety requirements.

\section{Experiment}\label{result}

\subsection{Intended behavior Model Comparison}
Mattas et al. \cite{mattas_driver_2022} have compared different intended behavior models using the HighD dataset. However, no study has yet compared these models using a Chinese traffic dataset. Yu et al. \cite{yu2022time} highlight significant differences between traffic flow characteristics on highways in China and Germany, which may influence the behavior of AV. To address this gap, we will conduct a comparison of intended behavior models using the AD4CHE dataset.

Cut-in, deceleration, and cut-out are common scenarios on highways \cite{tenbrock2021conscend}. Particularly, cut-in scenarios challenge both the longitudinal and lateral capabilities of an AV. To evaluate the safety of different intended behavior models, we measure their performance in response to cut-in scenarios.

\begin{figure}[t]
	\centering
        \includegraphics[width=0.45\textwidth]{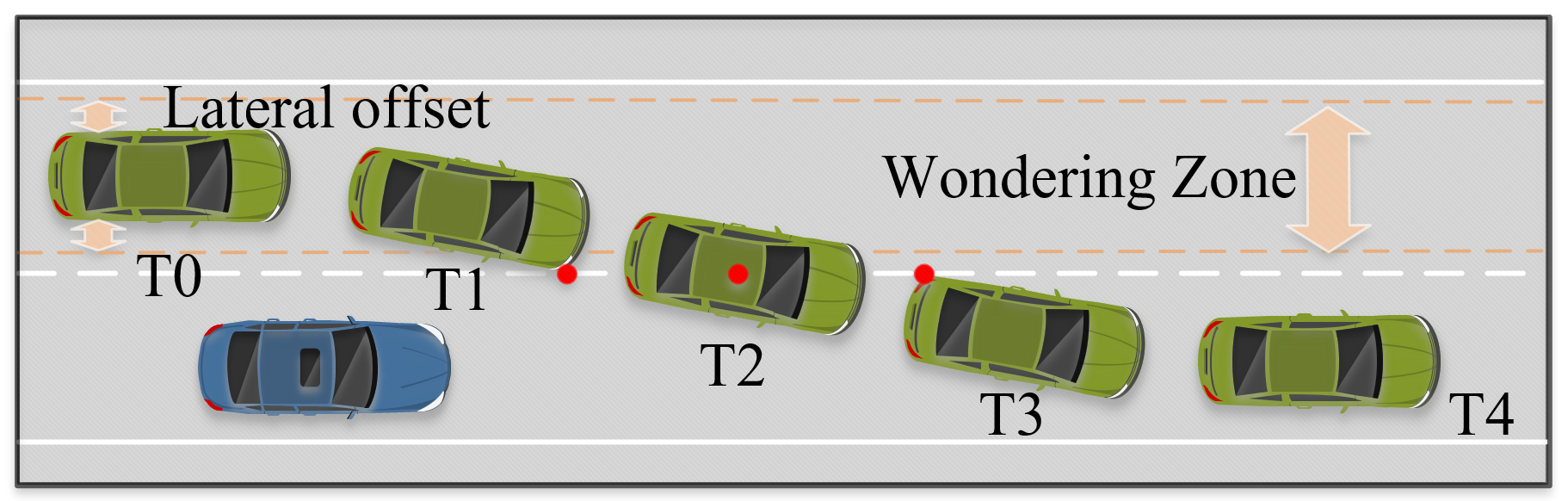}
	\caption{The definition of a cut-in scenario using four key timestamps. Red points represent key parts of the cut-in vehicle. The range of lateral offset, when a vehicle keeps a lane, is defined as a wandering zone.}
	\label{FIG:5}
\end{figure}

The dataset includes 53,761 vehicles, from which we extracted all cut-in vehicle pairs based on the defined criteria for cut-in, as illustrated in \cref{FIG:5}. Specifically, four key timestamps are defined. The cut-in vehicle exits the wandering zone at $T_0$, where the wandering zone refers to the range of lateral offset a vehicle maintains within a lane, which can be defined, for example, as 0.375 meters, as referenced in UN Regulation 157 \cite{un_ece_regulation_2021}. At $T_1$, the front right tire of the cut-in vehicle crosses the lane marking; at $T_2$, the vehicle's center crosses the lane marking; and at $T_3$, the rear left tire crosses the lane marking. Finally, $T_4$ marks the point when the cut-in vehicle re-enters the wandering zone. For scenarios where the vehicle cuts in from right to left, the definitions of $T_1$ and $T_3$ are adjusted accordingly.

Initially, we identified 3,026 cut-in scenarios. These were further filtered based on two criteria:
\begin{itemize} 
\item Cut-in scenarios where the TTC consistently exceeds 5 seconds are disregarded, as the cut-in vehicles in such cases are likely irrelevant to the ego vehicle's decision-making process;
\item Cut-in scenarios involving non-car traffic participants are excluded, as certain intended behavior models, such as FSM, have not yet been validated for these non-car traffic scenarios.
\end{itemize} 

After applying these filters, we obtained 334 valid cut-in scenarios. By applying the intended behavior models to these scenarios, we assessed their ability to prevent collisions. The parameters used for each model are detailed in \cref{table2}. Through simulation, we found that the CC Driver model resulted in collisions in three cut-in scenarios, while FSM and RSS models showed no collisions in any of the valid scenarios. Specifically, FSM and RSS models performed well because no actual collisions occurred in the dataset. The CC Driver model, however, requires further refinement to produce results consistent with real-world outcomes.

\begin{figure}[ht]
	\centering
    \subfloat[]{
        \includegraphics[width=0.8\linewidth]{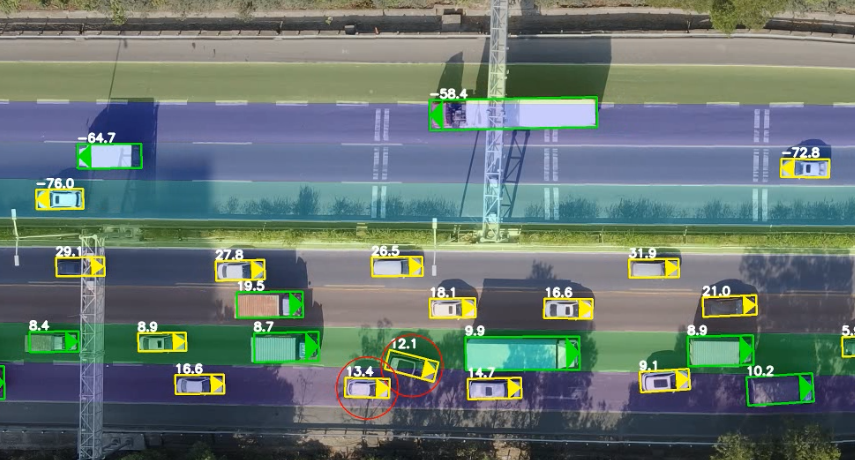}
        \label{FIG:6a}
    }
    
	\subfloat[]{
        \includegraphics[width=0.9\linewidth]{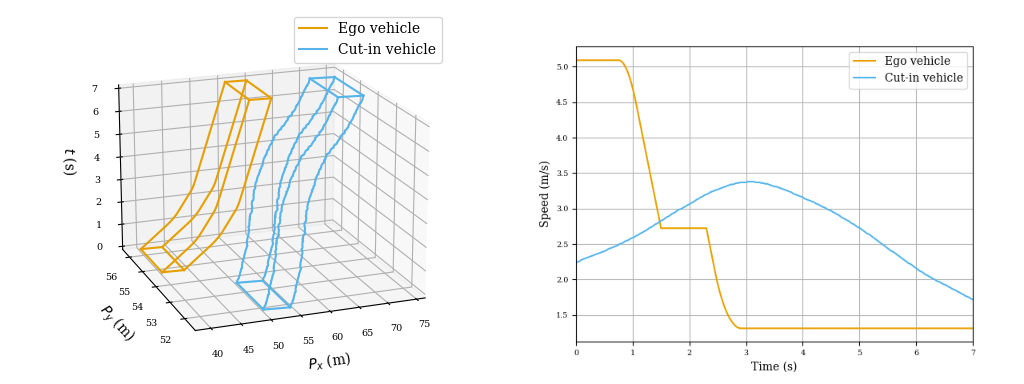}
        \label{FIG:6b}
    }
    
	\subfloat[]{
        \includegraphics[width=0.9\linewidth]{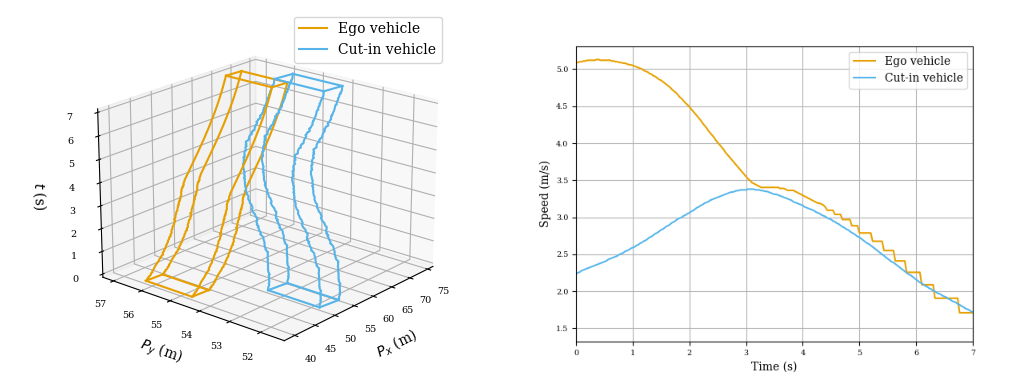}
        \label{FIG:6c}
    }
	\caption{Results of FSM and CC Driver models in one cut-in scenario. (a) The origin cut-in scenario in the dataset; (b) Position and velocity of both vehicles using the FSM model; (c) Position and velocity of both vehicles using the CC Driver model.}
	\label{FIG:6}
\end{figure}

Since both the RSS and FSM models successfully prevent collisions, we selected one of them to compare with the CC Driver model in a specific cut-in scenario to illustrate our findings, as shown in \cref{FIG:6a}. In this scenario, the ego vehicle initially travels at a higher speed than the cut-in vehicle. When the cut-in vehicle merges into the ego vehicle's lane and the safety checker detects a risk, the ego vehicle controlled by the FSM model decelerates much more sharply compared to the vehicle controlled by the CC Driver model. This difference in deceleration is evident when comparing the speed reductions in \cref{FIG:6b} and \cref{FIG:6c}. As a result, the CC Driver model fails to prevent a collision in time, leading to overlapping positions of the two vehicles at several time stamps. In contrast, no such overlap occurs in the FSM model, indicating its effectiveness in collision avoidance. Additionally, the FSM model allows the ego vehicle to adjust its response based on the level of risk, as observed in \cref{FIG:6b}, where the ego vehicle exhibits variable braking intensity.

\begin{table}[t]
\caption{Parameter values used in models for simulations.}
\centering
\label{table2}
\begin{tabularx}{0.4\textwidth}{XXX}
\hline
Models & Parameter & Value \\
\hline
\multirow{6}{*}{RSS} & $\rho$ & 0.75 $\rm s$ \\
~ & $a_{\rm max,accel}$ & 3 $\rm{m/s}^2$ \\
~ & $a_{\rm min,brake}$ & 6 $\rm{m/s}^2$ \\
~ & $a_{\rm max,brake}$ & 6 $\rm{m/s}^2$ \\
~ & $J_{\rm max}$ & 12.65 $\rm{m/s}^3$ \\
~ & $a_{\rm max}$ & 0.774 $\rm g$ \\
\hline
\multirow{6}{*}{FSM} & $\tau$ & 0.75 s \\
~ & $b_{\rm ego,comf}$ & 3 $\rm{m/s}^2$ \\
~ & $b_{\rm ego,max}$ & 6 $\rm{m/s}^2$ \\
~ & $b_{\rm cut-in,max}$ & 7 $\rm{m/s}^2$ \\
~ & $J_{\rm max}$ & 12.65 $\rm{m/s}^3$ \\
~ & $a_{\rm max}$ & 0.774 g \\
\hline
\end{tabularx}
\end{table}

While both FSM and RSS models achieve the desired outcome of collision avoidance in the studied scenarios, the RSS model has broader applicability. The FSM model struggles to identify risks when two vehicles maintain the same longitudinal speed and reacts too late in overlap cut-in scenarios, where the rear of the cut-in vehicle is behind the front of the ego vehicle at time step $T_1$. Therefore, we chose the RSS model to demonstrate our method for deriving subsystem-level SOTIF perception requirements. However, the FSM model still shows great potential as a viable intended behavior model. The calibration method proposed by Xu et al.\cite{xu_calibration_2021} could be beneficial in fine-tuning the FSM model. Nevertheless, calibration is beyond the scope of this paper.

\subsection{Results of the Collision Severity Model}
\begin{figure}[t]
	\centering
        \includegraphics[width=0.50\textwidth]{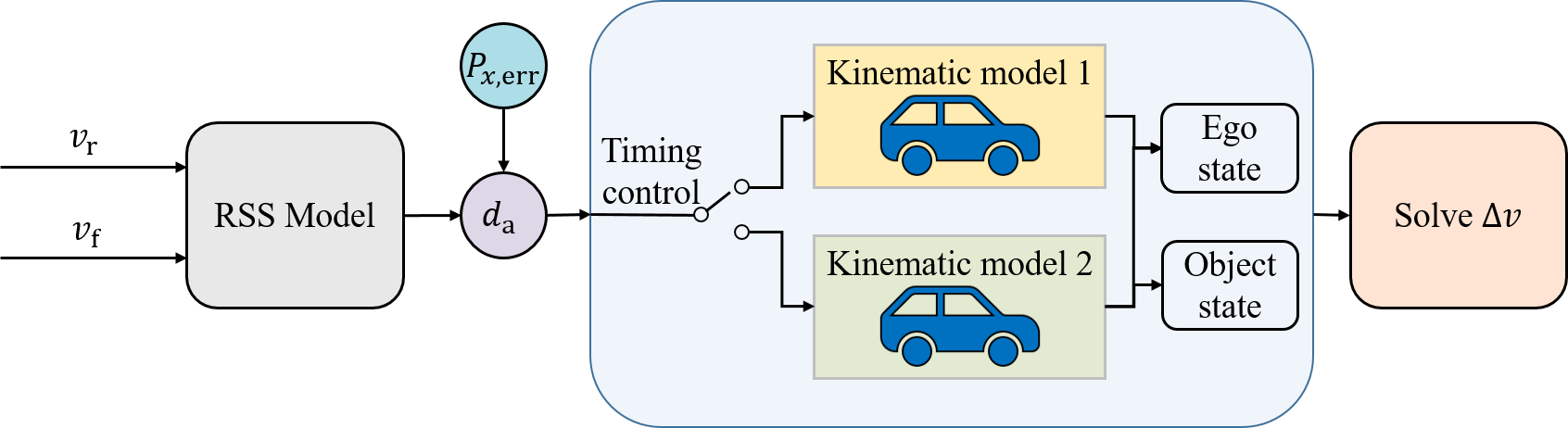}
	\caption{The process to solve $\Delta v$ at the collision time stamp given a specific position error. Depending on the timing, two different kinematic models are employed to update the states of the ego and the object. The $\Delta v$ is solved when a collision occurs via simulations.}
	\label{FIG:7}
\end{figure}

The model between state uncertainty and collision severity is established based on the theory in \cref{meth}. We selected the deceleration scenario defined within the RSS model: the leading vehicle brakes at its maximum deceleration rate, and after a designated reaction time, the following vehicle decelerates at its minimum rate. This scenario simulates subsystem-level safety requirements. In the RSS model, the minimum safe distance $d_{\rm min}$ is determined by $v_{\rm r}$ and $v_{\rm f}$. Injecting position errors $P_{\rm x,err}$ into $d_{\rm min}$ simulates the actual relative distance $d_a$ between vehicles, where $d_{\rm min}$ is the perceived safe distance, and $d_a$ is the actual distance. The AV does not adjust its driving policy, assuming a safe distance, but the actual distance is smaller, resulting in a velocity difference $\Delta v$ before the collision. \cref{FIG:7} shows the process of solving $\Delta v$ for a given position error. The rear and front vehicle states are iteratively updated until $d_a$ becomes negative, with $\Delta v$ at this moment being the solution. During iterations, three key time stamps are used to apply two different kinematic models for state updates:
\begin{itemize}
    \item 	$t_1$: the time stamp when the reaction time $\rho$ is reached;
    \item 	$t_2$: the time stamp when the rear vehicle brakes to a standstill;
    \item 	$t_3$: the time stamp when the front vehicle brakes to a standstill.
\end{itemize}

\cref{eq11} and \cref{eq12} are utilized to update the states of the rear vehicle in each time step of $\Delta t$. If $t\le t_1$, the rear vehicle is still in the acceleration phase due to the reaction time, whose states are determined by the kinematics model 1.  After $t_1$, the rear vehicle begins to decelerate, and its states are determined by the kinematic model 2. If $t>t_2$, the rear vehicle already stops, and its state remains the same as the previous states. 
\begin{flalign} \label{eq11} 
\begin{split}
P_{\rm x,t,r}&=P_{\rm x,t-1,r}+
\\&
\begin{cases}
    v_r\Delta t+\frac{1}{2}a_{\rm max,accel}\Delta t^2,\ &\text{if}\  t\leq t_1\\
    v_r\Delta t-\frac{1}{2}a_{\rm min,brake}\Delta t^2,\ &\text{if}\  t_1<t\leq t_2
\end{cases}
\end{split}&
\end{flalign}

\begin{equation}  \label{eq12}
v_{\rm x,t,r}=v_{\rm x,t-1,r}+\begin{cases}
    a_{\rm max,accel}\Delta t,\ &   \text{if}\  t\leq t_1\\
    -a_{\rm min,brake}\Delta t,\ &\text{if}\  t_1<t\leq t_2
\end{cases}\end{equation}

The states of the front vehicle are updated by \cref{eq13} and \cref{eq14}: If $t\le t_3$, the front vehicle decelerates continuously. Thus, the kinematic model 2 is used to update its states. After $t_3$, the state of the front vehicle remains unchanged.
\begin{flalign}
P_{\rm x,t,f}=P_{\rm x,t-1,f}&+v_f\Delta t-\frac{1}{2}a_{\rm max,brake}\Delta t^2 ,\ \text{if} \ t\leq t_3& \label{eq13}
\\
    v_{x,f}=&v_{0,f}-a_{\rm max,brake}t,\ \text{if}\ t\le t_3&\label{eq14}
\end{flalign}

After updating the states of the rear and front vehicles in each iteration, the actual relative distance is also recalculated until a negative value of $d_a$ is reached. Specifically, the simulation ends when a collision occurs. Finally, the differentiated velocity $\Delta t$ is obtained at the time stamp of a crash.
\begin{figure}[t]
	\centering
        \includegraphics[width=0.40\textwidth]{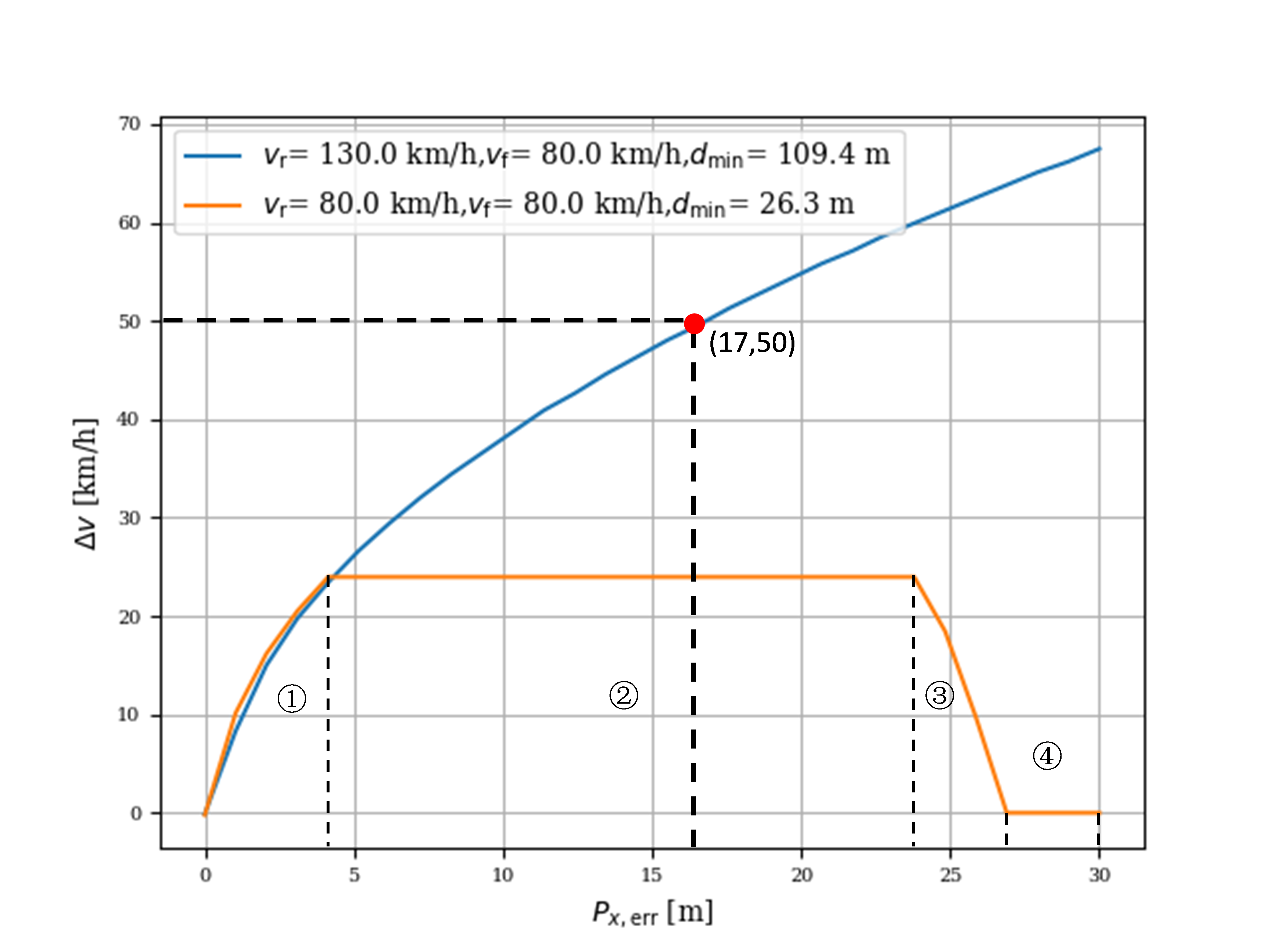}
	\caption{The relation between position errors and collision severity using the RSS model. The red point illustrates the maximum allowable longitudinal position error given an acceptable collision severity. \ding{172} represents that the front vehicle brakes until standstill; \ding{173} represents that the front vehicle is not still, Due to the same deceleration rate, $\Delta v$ remains constant; \ding{174} represents that the rear vehicle still in the acceleration phase, and the front vehicle is not still; \ding{175} is invalid because the position error is larger than the actual distance.}
	\label{FIG:8}
\end{figure}

We employ the RSS model as the driving policy for an AV when its distance from another vehicle is less than the minimum safe distance. While other driving policies could be used, this paper specifically demonstrates how safety requirements can be derived based on the selected RSS model. Following the process outlined in \cref{FIG:7}, we establish the relationship between position errors and collision severity, as depicted in \cref{FIG:8}. Since our focus is on highway scenarios, we consider two cases with different minimum safe distances ($d_{\rm min}$).

In the first case, the rear vehicle travels at 130 km/h, while the front vehicle moves at 80 km/h. The significant speed difference necessitates a large $d_{\rm min}$ of 109.4 meters. As position errors increase, the rear vehicle has less distance to decelerate, resulting in a higher $\Delta v$ upon collision. In contrast, the second case involves both vehicles traveling at 80 km/h, requiring a smaller $d_{\rm min}$ of 26.3 meters. This smaller $d_{\rm min}$ produces different results, as shown by the yellow line in \cref{FIG:8}.

The yellow line is divided into four segments. In segment \ding{172}, the rear vehicle behaves similarly to case one. In segment \ding{173}, the front vehicle is still decelerating before the collision, leading to a constant $\Delta v$ due to the same deceleration rate. In segment \ding{174}, significant position errors cause the rear vehicle to accelerate, resulting in an earlier collision and a decreasing $\Delta v$. Finally, in segment \ding{175}, the position error becomes so large that the rear vehicle no longer needs to react to the front vehicle.

Given a maximum acceptable collision severity, defined as $\Delta v_{\rm max}$ of 50 km/h \cite{buerkle_safe_2021}, the maximum allowable position error is determined to be 17 meters, as indicated by the red point in \cref{FIG:8}. Since case one involves a relatively large velocity difference, the maximum allowable position errors derived from other possible velocity combinations in highway scenarios are smaller than those in case one. Therefore, the results from case one are considered the requirements for position accuracy in object tracking.
\begin{figure}[t]
	\centering
        \includegraphics[width=0.4\textwidth]{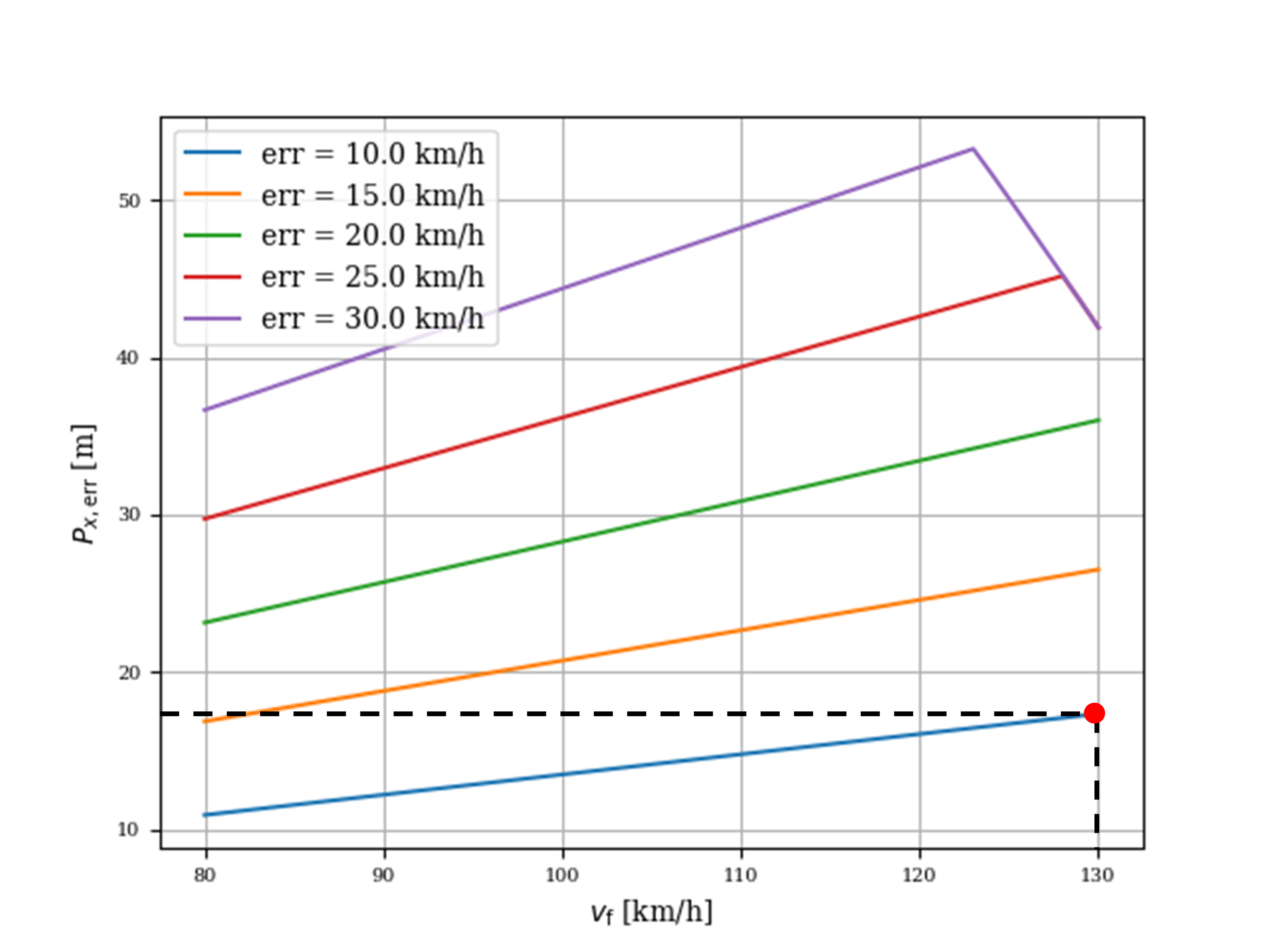}
	\caption{The relation between the velocity error and position error with a speed of 130 km/h for the rear vehicle. Given a derived maximum allowable longitudinal position error, the maximum acceptable velocity error is obtained, as indicated by the red point. }
	\label{FIG:9}
\end{figure}

Except for the requirements for position accuracy, velocity accuracy is also important to evaluate object tracking performance. Instead of directly correlating velocity errors with collision severity, we examine how velocity errors impact position errors. \cref{FIG:9} illustrates the possible velocities of the front vehicle in highway scenarios, assuming a constant speed of 130 km/h for the rear vehicle. As depicted in the figure, position error increases as velocity error grows. Given that the requirement for position accuracy has already been established at 17 meters ($P_{\rm x,err}$), we determine that the maximum allowable velocity error is approximately 10 km/h. This threshold ensures that the position error does not exceed the maximum permissible limit, regardless of the front vehicle's velocity. Therefore, to prevent severe collisions, the velocity error must be kept below 10 km/h.
\subsection{Results of the Bayesian Model}
Unlike the model used for deriving requirements for state uncertainty, a Bayesian model to link system-level risk and existence uncertainty is applied, as expressed in \cref{eq9}. We continue to employ the RSS model and its corresponding deceleration scenario for our simulation experiments in the HighD dataset. Low collision risk will emerge as FP objects will be responded to according to the RSS model. As a result, FP objects are not considered in this experiment. Therefore, we only calculate the FN rate to demonstrate the proposed model. Compared to \cref{eq9}, $N$ is reduced to 1, representing the FN only. Given a specific value of $p_{\rm C|E}$ and $p_{\rm S|C}$, \cref{eq9} can be rewritten as:
\begin{flalign}\label{eq15}
&&
    \lambda^\prime=\sum_{i=1}^{M}{{(p}_{{\rm PL},i}\times p_{{\rm E|PL},{i}})}
    &&
\end{flalign}
where $M$ is the number of distance partitions in front of the ego vehicle. $\lambda^\prime$ is the equivalent acceptable system-level risk. 



\begin{figure}[t]
	\centering
        \includegraphics[width=0.45\textwidth]{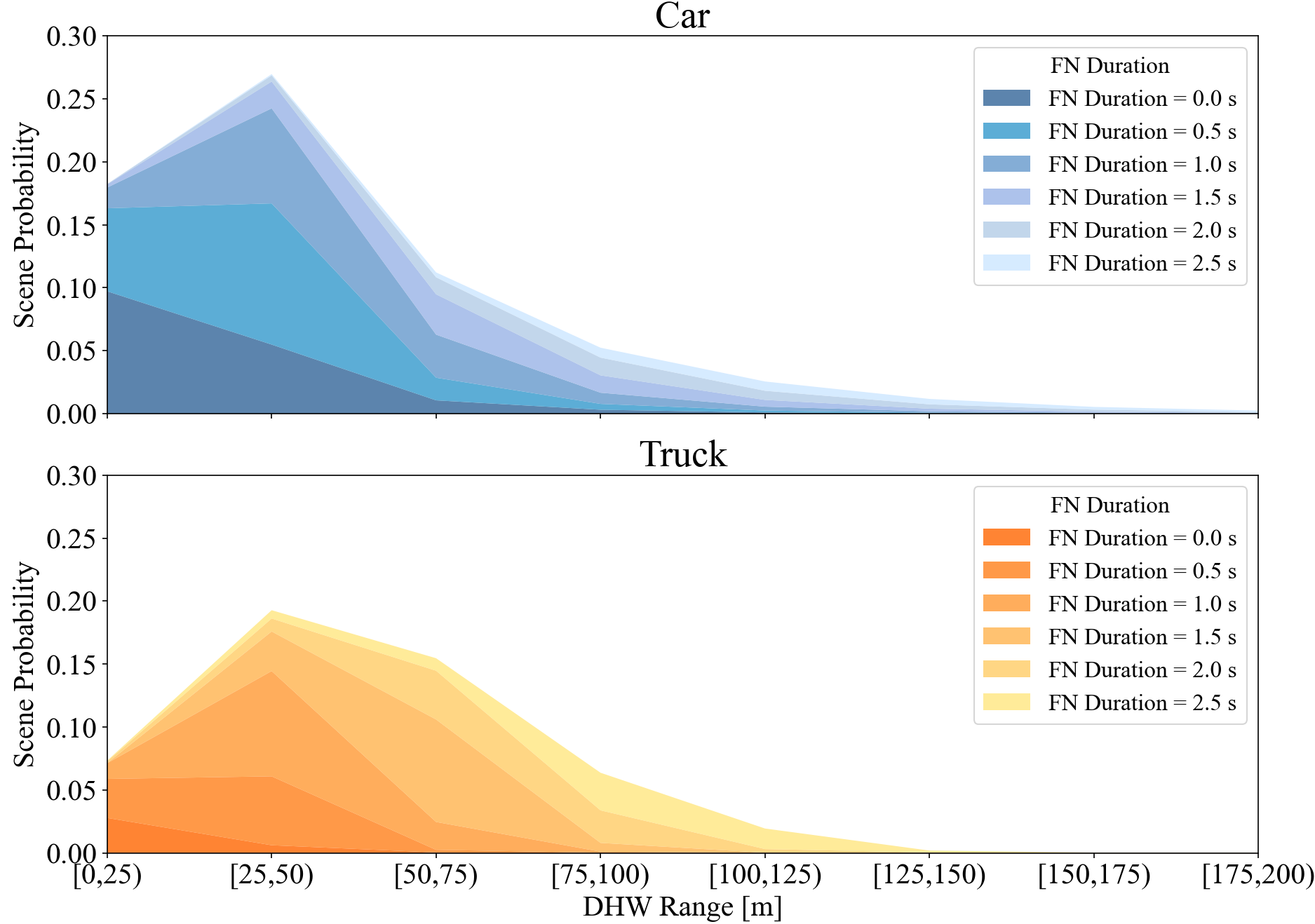}
	\caption{The probability of relevant scenarios in different distance partitions considering the duration of an FN object. }
	\label{FIG:11}
\end{figure}

The implementation of the model begins by identifying vehicle pairs in the dataset that do not involve lane changes. The data is first divided into two categories based on the type of rear vehicle. Their scenario parameters are then used as the initial parameters for the RSS model, with uncertainty injected into the rear vehicle from high to low. All scenarios resulting in collisions are recorded as relevant scenarios. By analyzing the distribution of $d_a$ within these scenarios, the relative occurrence probability of relevant scenarios $p\left\{d(S)\in A_i\right\}$ is calculated for each predefined distance partition $A_i$. The final occurrence probability of relevant scenarios for the $i$-th distance partition under the $j$-th PL and $k$-th module, $p_{{\rm E|PL},{i,j,k}}$, is determined by the ratio of relevant scenarios to the total number of scenarios in the dataset. The result is shown in \cref{FIG:11}.

The darkest block in the graph represents the probability of the relevant scenario when ``FN Duration = 0.0 s", which is equivalent to not injecting uncertainty. The reason is that the rear vehicle is already engaged in a scenario that the RSS model cannot manage. Furthermore, \cref{FIG:11} highlights a significant trend: the probability of relevant scenarios is markedly higher within near-distance partitions, especially within the distance headway (DHW) range of [25, 50) meters. This finding emphasizes the importance of focusing on FN objects within near-distance partitions when evaluating system-level risk. Practically, if we assume a uniform cost for reducing FN risks across different distance partitions, prioritizing mitigation efforts within the [25, 50) meter range would be more effective and efficient.

Based on the acceptable system-level risk and the derived probability of relevant scenarios for FN objects, the acceptable FN rate $p_{\rm PL}$ for different distance partitions can be determined as outlined in \cref{eq15}. Given that the probability of relevant scenarios is higher in near-distance partitions, it is logical to assign a lower acceptable FN rate to these partitions compared to those farther away, since $\lambda^\prime$ remains constant. 

The assignment of specific safety requirements can be viewed as an optimization problem:

\begin{equation}
\begin{split}
\label{eq:opti1}
\min_{P} \quad & f_{\text{cost}}(P) \\
\text{s.t.} \quad & P \in \{0,1\}, \\
&  P \cdot E \leq \lambda
\end{split}
\end{equation}
\( P \) is a one-dimensional increasing array composed of the safety requirements \( p_{PL,i} \) to be allocated. \( E \) is the related scenario probability \( p_{{\rm(E|PL)},i} \) , \( i \) is the number of DHW partitions. \( \lambda \) is the system-level safety requirement to be decomposed, calculated using the MEM safety principle, which is $\rm 6.8 \times 10^{-7}/h$ based on US accident rate\cite{berk_safety_2019}. \( f_{\text {cost }}(P)=\sum_{i=1}^{8} \frac{\ln (1-C)}{\ln \left(1-p_{P L, i}\right)} \)  is the cost function,  which means minimizing the test mileage of the allocated safety requirements\cite{kalra_driving_2016}. Taking the scenario probability of ``FN Duration = 1 s" as an example, the calculation results are shown in \cref{tab:sys_result}.

\begin{table}[t]
    \centering
    \caption{subsystem-level Safety Metric Calculation Results}
    \label{tab:sys_result}
    \begin{tabular}{m{3.5cm}m{3.5cm}}
    \hline
\textbf{Distance Range (m)} & \textbf{Safety Requirements (/h)} \\
\hline
         
        $[0, 25)$            & $5.08 \times 10^{-6}$  \\ 
        $[25, 50) $          & $5.87 \times 10^{-6}$  \\ 
        $[50, 75)$           & $2.63 \times 10^{-5}$  \\ 
        $[75, 100)$          & $8.23 \times 10^{-5} $ \\ 
        $[100, 125)$         & $2.07 \times 10^{-4}$  \\ 
        $[125, 150)$         & $4.65 \times 10^{-4} $ \\ 
        $[150, 175)$         & $7.48 \times 10^{-4}$  \\ 
        $[175, 200) $        & $1.56 \times 10^{-3}$  \\ 
        \hline
    \end{tabular}
\end{table}


\subsection{Experimental design and analysis of results of component-level method}
We evaluate the EagerMOT algorithm\cite{kim2021eagermot}, a MOT method based on a greedy strategy. EagerMOT tracks targets by selecting the best-matching detection for the current trajectory, combined with trajectory prediction to improve accuracy and efficiency. As a traditional data fusion method, it allows flexible replacement of input detection results, simplifying debugging.  In experiments, TrackRCNN\cite{voigtlaender2019mots} served as the 2D detector and PointGNN\cite{shi2020point} as the 3D detector. Their detection results on the KITTI MOT dataset were inputs for EagerMOT. The first 15 sequences formed the training set, and the subsequent 6 sequences were used for testing.



The next step is to build the perturbation metric dataset. But we found that using 0 as background data for random replacement was unfeasible, as it totally disrupted system operation. Instead, detection ground truth was chosen as background data. For example, to replace the FN rate of the 2D bounding box: original camera data were evaluated to obtain metric data, and perturbations were added to generate perturbed metrics. FN data were extracted by comparing true values with original data, then added proportionally to generate perturbed data. 

\begin{figure}[htbp]

\centering

\includegraphics[width=0.45\textwidth]{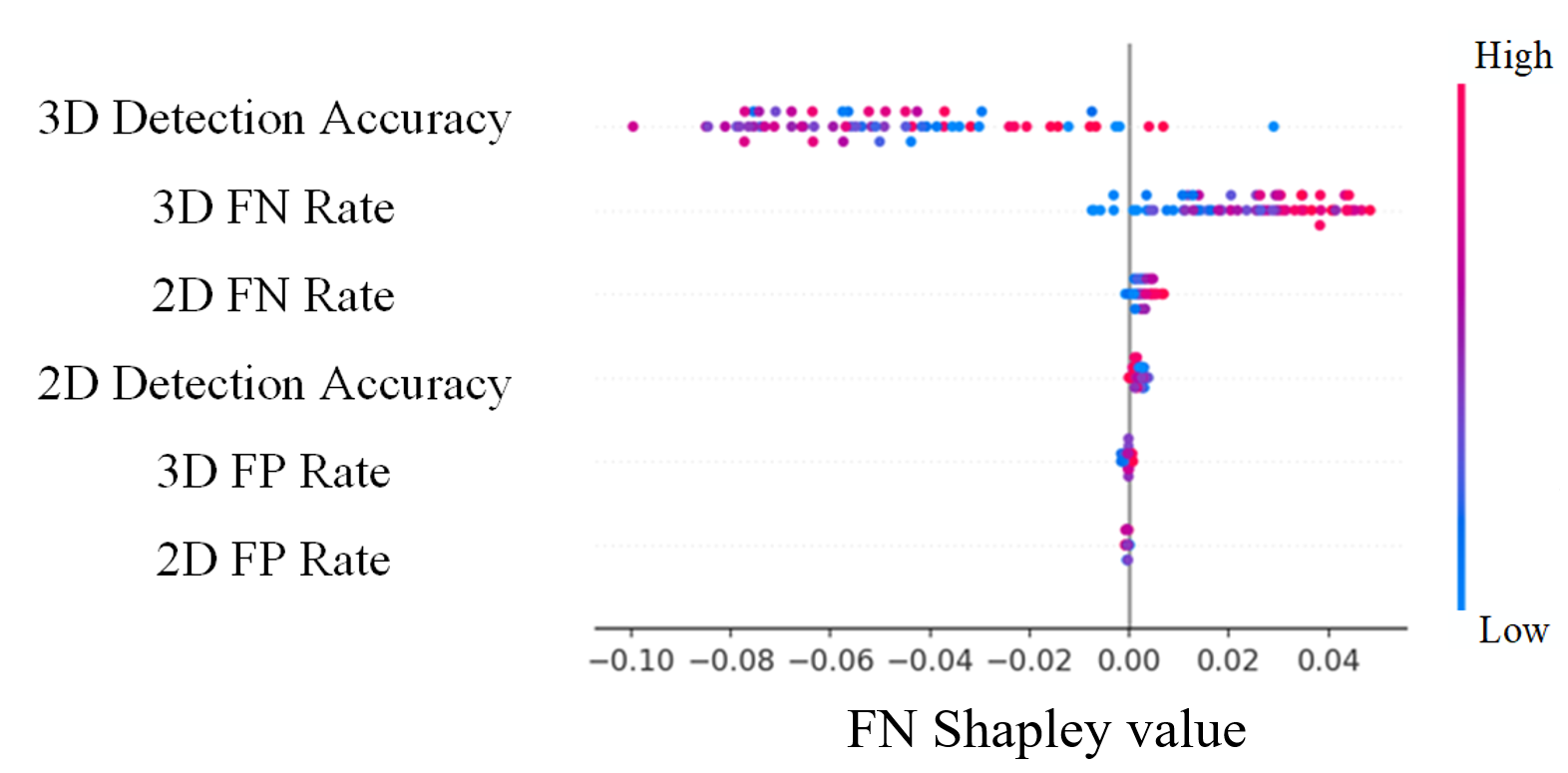}

\caption{Shapley values of each input feature on the FN metric. }

\label{fig:shapley-values}

\end{figure}

Perturbed detection data was used as input for the MOT algorithm, whose output performance derived perturbed metric data. We have obtained a metric dataset of input data and corresponding output data for the MOT algorithm, which will be used for training the SHAP model. The model will output Shapley values which helps us understand the performance changes of MOT systems when facing input data of different qualities. To ensure numerical stability, all input and output metric data were normalized. For detection accuracy metrics, the average normalized value was used as a single 2D/3D detection metric. Perturbation metric data were generated in steps of 0.01 from 0 to 1, creating 101 groups of data. After training the model with these data, the Shapley value results are presented using the FN metric as an example, as shown in \cref{fig:shapley-values}.

The visualization comprises multiple points, with each point representing three features: vertical position indicates the depicted feature, color denotes whether the feature value is high or low, and horizontal position signifies whether the value's influence leads to higher or lower predictions. Input features within each graph are arranged from top to bottom based on the absolute mean value of their Shapley values. The average Shapley values of each input feature are used as a measure of feature importance. After aggregating the data across all evaluation metrics, including the FN metric, the results are presented in \cref{tab:metrics_results}.

\begin{table*}[ht]
\centering
\caption{Shapley value calculation result. The data in the table represents the Shapley value of the component-level safety requirement in the column compared to the subsystem-level safety requirement in the row.}
\label{tab:metrics_results}
\begin{tabular}{lcccccc}
\hline
\textbf{Metric} & \textbf{2D Detection Accuracy} & \textbf{2D FN Rate} & \textbf{2D FP Rate} & \textbf{3D Detection Accuracy} & \textbf{3D FN Rate} & \textbf{3D FP Rate} \\
\hline
FN   & -0.00189 & 0.00300 & 0.00043 & -0.04758 & 0.02310 & -0.00028 \\
FP   & -0.00040 & -0.00102 & 0.00502 & -0.00524 & 0.00212 & 0.02316 \\
AssA & 0.00282 & -0.00593 & -0.00190 & 0.02597 & -0.02443 & -0.00986 \\
LocA & 0.00022 & -0.00034 & -0.00003 & 0.09112 & -0.00682 & -0.00050 \\
DetA & 0.00139 & -0.00197 & -0.00390 & 0.04513 & -0.02355 & -0.02077 \\
HOTA & 0.00210 & -0.00394 & -0.00291 & 0.03558 & -0.02400 & -0.01534 \\
\hline
\end{tabular}
\end{table*}

Subsequently, based on the type of subsystem-level safety requirements, MOT evaluation metrics are selected, and the metrics for each component are summed. Quantitative component-level safety requirements are then allocated based on the measured impact. For instance, a subsystem-level safety requirement like ``the FN rate of vehicles within 25 to 50 meters ahead should not exceed \( 5.87 \times 10^{-6} \)" translates to a requirement on the MOT system's FN rate. This can be modeled as an optimization process using data in \cref{tab:metrics_results} :
\begin{equation}
\begin{split}
\label{eq:opti}
\min_{Z} \quad & f_{\text{cost}}(Z) \\
\text{s.t.} \quad & Z \in \{0,1\}, \\
&  Z \cdot \phi + \phi_0 \leq 5.87 \times 10^{-6}
\end{split}
\end{equation}
where \( Z = [z_1, z_2, z_3, z_4, z_5, z_6] \) is a one-dimensional vector representing the allocated component-level safety requirements, where each element corresponds to 2D detection accuracy, 2D FN rate, 2D FP rate, 3D detection accuracy, 3D FN rate, and 3D FP rate. \( \phi = [\phi_1, \phi_2, \phi_3, \phi_4, \phi_5, \phi_6] \) is also a one-dimensional vector representing the feature importance of each element in \( Z \) on the MOT FN rate, matching the data in the first row of \cref{tab:metrics_results}. \( \phi_0 \) denotes the baseline value, signifying the FN rate of the MOT algorithm when applied to the unoptimized original dataset, which was 0.048 in this research. \( f_{\text{cost}} \) is the cost function representing the total cost of optimizing the input features, which is assumed to be infinitely large if the values of the features approach their ground truth, i.e., \( f_{\text{cost}} = \sum_{i=1}^{6}\frac{1}{1-z_i} \) considering practical development constraints. The optimization results are detailed in \cref{tab:decom_results} , using the \textit{fmincon} solver from the MATLAB Optimization Toolbox\cite{matlab_optimization_toolbox}.

\begin{table}[t]
\centering
\caption{Detection Accuracy and False Positive/Negative Rates}
\label{tab:decom_results}
\begin{tabular}{ccc}
\hline
\textbf{2D Detection Accuracy} & \textbf{2D FN Rate} & \textbf{2D FP Rate}  \\ 
\hline
87.8\% & 9.67\% & 74.45\%  \\ 
\hline
\textbf{3D Detection Accuracy} & \textbf{3D FN Rate} & \textbf{3D FP Rate}\\
\hline
97.6\% & 3.49\% & 68.34\%\\
\hline
\end{tabular}
\end{table}

For the safety requirements at the component level assigned, verification is performed from two aspects of performance. The first aspect is efficiency verification, which examines whether component-level safety requirements allocated for subsystem-level failures provide optimal performance optimization for those subsystem-level failures. The second aspect is effectiveness verification, which assesses whether the safety requirements allocated to the component level ensure that the safety requirements at the subsystem level are met.

\begin{figure}[ht]
    \centering
    \includegraphics[width=\linewidth]{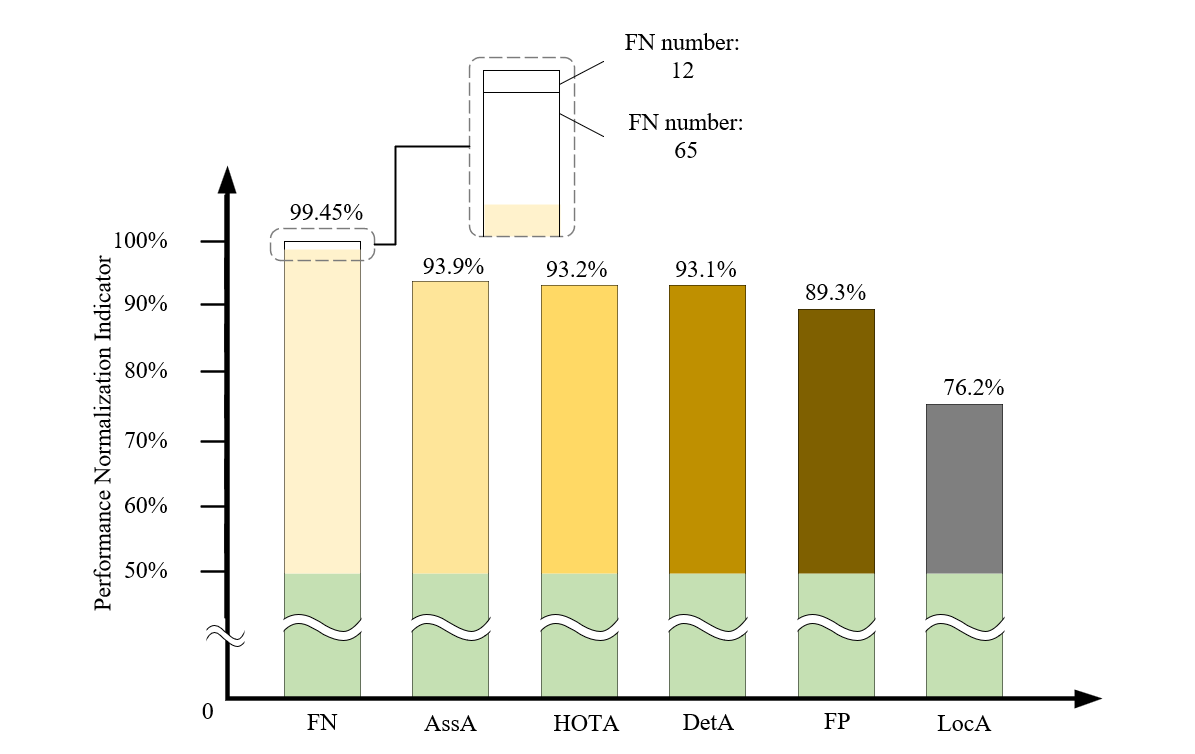}
    \caption{Algorithm performance statistics after meeting component-level safety requirements}
    \label{fig:algorithm_performance}
\end{figure}

In the verification experiment, the test set data are perturbed according to the component-level safety requirements to obtain perturbed data. The perturbed data are then used as input to the MOT algorithm, and the results are quantitatively evaluated to obtain the perturbation metrics. The degree to which the perturbation metrics meet the subsystem-level safety requirements is observed, as shown in \cref{fig:algorithm_performance}.

To unify the dimensions, the perturbation metrics are normalized. The MOT algorithm's output metrics with ground truth data as input are set to 100\%, while the output metrics with original data as input are set to 50\%. \cref{fig:algorithm_performance} indicates that after meeting the allocated component-level safety requirements, the EagerMOT algorithm's FN rate performance has reached 99.45\%, which is the most significant improvement among all output evaluation metrics, thereby completing the efficiency verification.

The experiment also shows that among the 1000 sets of safety metrics, only 65 sets had 1  FN frame, and 12 sets had 2 FN frames. Considering that the verification set has a cumulative total of 5793 frames, the ratio of the FN frames to the total frames across 1000 experiments is calculated, resulting in an FN rate of \(1.38 \times 10^{-4}\). This meets the subsystem-level safety requirements, thereby completing the effectiveness verification.

\section{ Discussion }\label{disc}
In this paper, we focus on the decomposition and quantification of SOTIF requirements using a specified top-level acceptance criterion. Our models effectively address subsystem-level safety requirements, particularly concerning state and existence uncertainties. Notably, both the collision severity and Bayesian models deliver promising results. The Bayesian model's use of distance partitions aligns well with sensor detection capabilities, enhancing its practical applicability. Unlike other studies, our deduced SOTIF requirements are both intuitive and explicit. For instance, a position error should not exceed 17 meters at a distance of approximately 110 meters when the ego vehicle travels at 130 km/h with a relative speed of 50 km/h to the leading vehicle.

To address component-level safety requirements, we propose a Shapley value-based method that decomposes the safety requirements of an AV perception system into those for individual components. Applying this method to a MOT system, we successfully derive SOTIF requirements for both camera-based and Lidar-based detection. This approach offers valuable guidance for system design and development. Finally, We randomly generated 1000 sets of safety requirements using the proposed method for simulation testing, and the experimental results proved the effectiveness and efficiency of the method.

However, while our focus has been on collision-related scenarios, it is important to acknowledge that other types of accidents, such as non-collision failures, were not considered in this study. Future research may explore these additional accident types to provide a more comprehensive safety assessment of autonomous vehicle systems.

In addition to this limitation, it is important to note that our verification experiment primarily establishes that the component-level safety requirements we derive serve as an upper bound for subsystem-level safety requirements. This suggests that these component-level requirements might be overly stringent. Future research could investigate the joint probability of failure across different components, such as camera-based and Lidar-based detection, which may be influenced by common factors, to refine these requirements.

Finally, given the use of a probabilistic model in our approach, acquiring more valid data is crucial for enhancing the reliability of our results. This highlights the need for accessible scenario datasets. As discussed in \cref{meth}, intended behavior models represent simplified versions of AV planning systems. For specific applications, developers should choose suitable datasets and employ their own planning systems, rather than relying on the model presented in this paper, to achieve more realistic design requirements.

\section {Conclusion and future work}\label{con}
This study proposes a SOTIF requirements decomposition method. To address subsystem-level safety requirements, the Bayesian model and collision severity model are introduced. For component-level safety requirements, a decomposition method based on the Shapley value is introduced, clarifying the safety requirements of ADS in different scenarios. Experimental results indicate that this method can effectively decompose system-level safety requirements into subsystem-level and component-level requirements, and the effectiveness of the component-level method is verified through practical application in the MOT system.

The methods proposed in this study achieved good results in experiments but still have some limitations. For example, the experimental data used in this study primarily comes from publicly available datasets, which may not fully represent the complex situations of real driving environments.

Nevertheless, the findings of this study provide new ideas and methods for the safety design of AV systems. By quantifying and decomposing subsystem-level and component-level safety requirements, this approach can better guide the development and testing of AV, thereby enhancing their safety and reliability in practical applications.

Future research can further refine the proposed methods, including extending this approach to planning and control systems and optimizing the computational efficiency of safety requirements decomposition. Additionally, incorporating more real-world scenario data for verification could improve the applicability and accuracy of the methods.

\section*{Funding information}
This work was supported by the National Natural Science Foundation of China (No. 52075213) and the Industrial Technology Basic Public Service Platform Project 2020 from the Ministry of Industry and Information Technology of China (No. 2020-0100-2-1). 


\bibliographystyle{IEEEtran}
\bibliography{reference}

\begin{IEEEbiography}[{\includegraphics[width=1in,height=1.25in,clip,keepaspectratio]{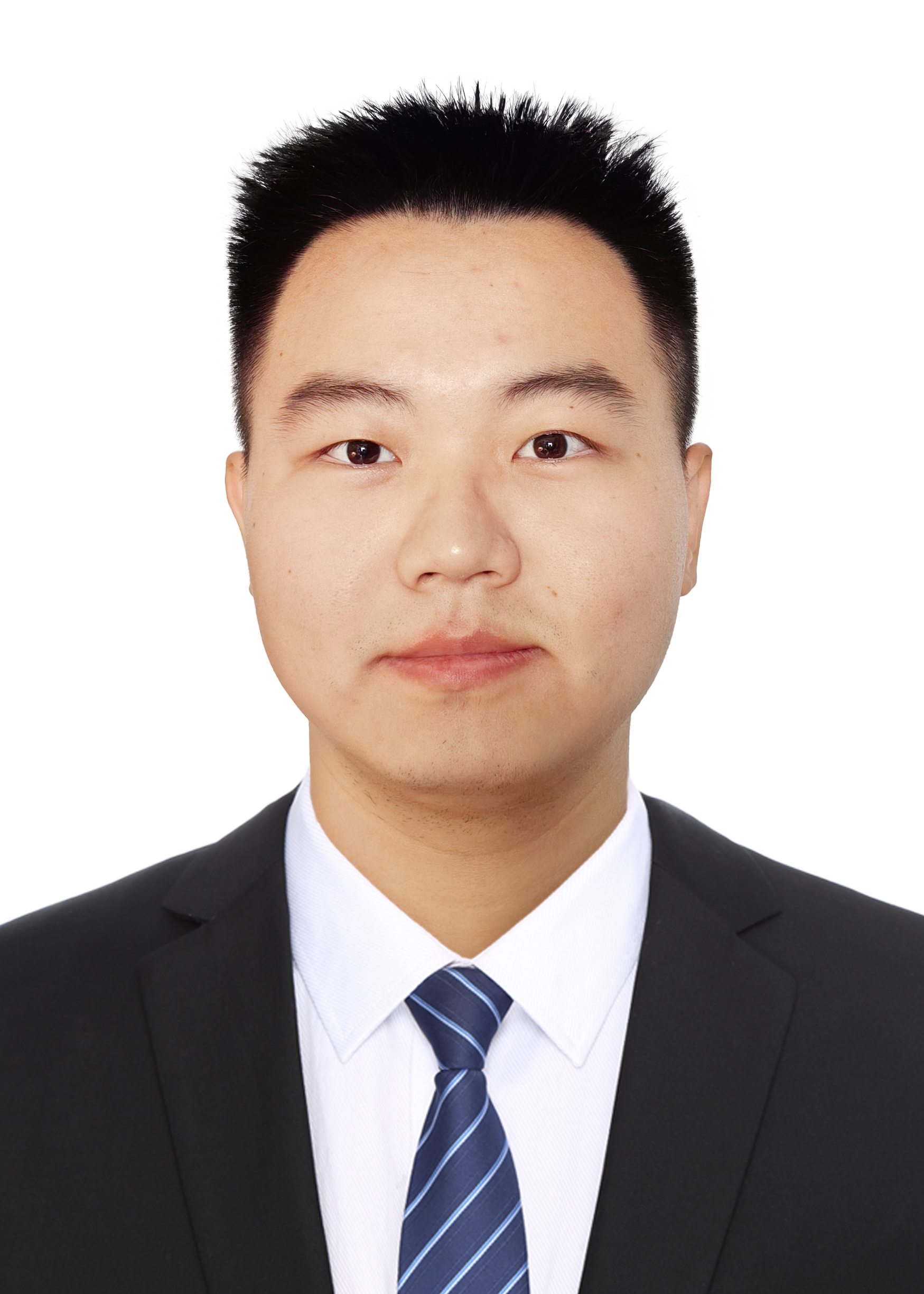}}]{Ruilin Yu}
received the B.S. degree in vehicle engineering from Hefei University of Technology, Hefei, China, in 2021, and the M.S. degree in vehicle engineering from Jilin University, Changchun, China, where he is currently pursuing the Ph.D. degree in vehicle engineering. His research interests include vehicle trajectory data analysis, safety of autonomous driving vehicles and perception system of automated vehicles.
\end{IEEEbiography}

\begin{IEEEbiography}[{\includegraphics[width=1in,height=1.25in,clip,keepaspectratio]{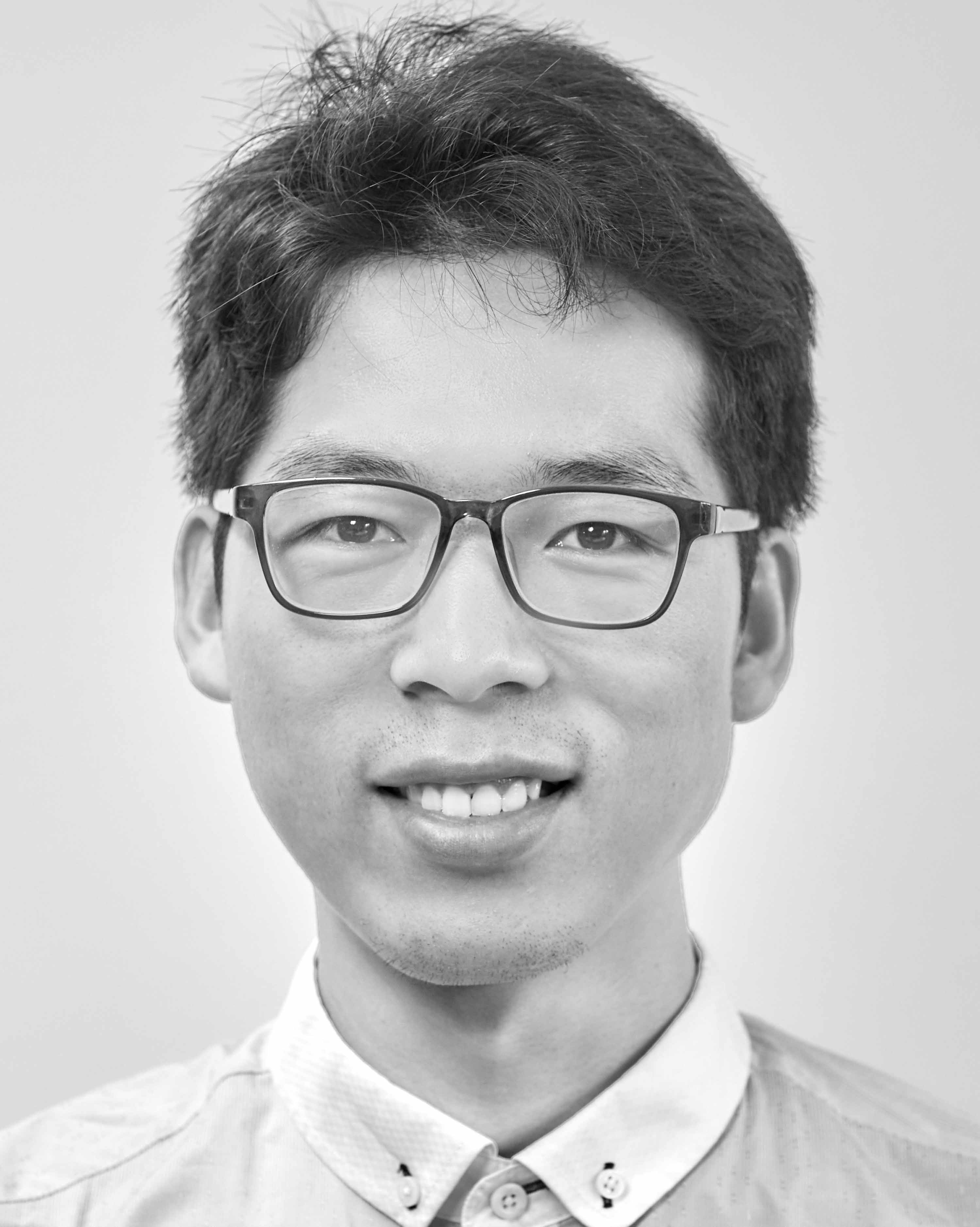}}]{Cheng Wang}
received the B.S. degree from the school of automotive engineering at Wuhan University of Technology, Wuhan, China, in 2014 and the M.S. degree from the school of automotive studies at Tongji University, Shanghai, China, in 2017 and the Ph.D. degree from the institute of automotive engineering at the Technical University of Darmstadt, Darmstadt, Germany, in 2021. From 2022 to 2024, he worked as a research associate at the University of Edinburgh. He is now an assistant professor at the Heriot-Watt University, Edinburgh. His research interest includes explainable AI and safety verification and validation of autonomous vehicles.
\end{IEEEbiography}

\begin{IEEEbiography}[{\includegraphics[width=1in,height=1.25in,clip,keepaspectratio]{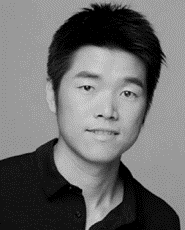}}]{Yuxin Zhang}
(Member IEEE) received the Joint Ph.D. degree in Vehicle Engineering from Jilin University, China and UC Berkeley, USA, in 2016. From 2016 to 2019, he worked as a postdoc researcher in Systems Engineering in Jilin University, China, and also a Safety Researcher in UISEE, Beijing, China. Since 2019, he worked as an Associate Professor at State Key Laboratory of Automotive Simulation and Control, Jilin University, China, and also as a Safety Researcher in DJI Automotive. His main research interests include automated driving systems safety engineering, functional safety, and safety of the intended functionality. 
\end{IEEEbiography}

\begin{IEEEbiography}[{\includegraphics[width=1in,height=1.25in,clip,keepaspectratio]{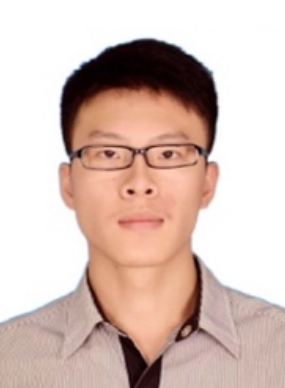}}]{Fumin Zhao}
received M.Sc. degree in mechanical and electronic engineering from Hefei University of Technology, Hefei, China, in 2017. Currently, he works in DJI Automotive. His research interests include SOTIF and Functional Safety of automated driving systems.
\end{IEEEbiography}

\end{document}